\documentclass[twocolumn,showpacs,amsmath,amssymb,prb,superscriptaddress]{revtex4}

\usepackage[pdftex]{graphicx} 
\usepackage{epstopdf}
\usepackage{verbatim}
\usepackage[T1]{fontenc}
\usepackage[latin1]{inputenc}
\usepackage[extra]{tipa}

\def\be{\begin{equation}}
\def\ee{\end{equation}}
\def\bea{\begin{eqnarray}}
\def\eea{\end{eqnarray}}
\def\ba{\begin{align}}
\def\ea{\end{align}}
\def\tp{\tilde{\phi}}

\def\HH{{\mathcal H}}
\def\HHlow{{\mathcal H}_\text{low}}

\def\hatppn{{\frac{\hat{\partial}}{\partial \phi_n}}}

\def\hatpn{\hat{\phi}_n}

\usepackage{color}


\begin{document}

\title{Universal transport signatures of Majorana fermions in superconductor-Luttinger liquid junctions}

\author{Lukasz Fidkowski}
\affiliation{Station Q, Microsoft Research, Santa Barbara, CA 93106-6105}
\author{Jason Alicea}
\affiliation{Department of Physics and Astronomy, University of California, Irvine, CA 92697}
\author{Netanel H. Lindner}
\affiliation{Institute of Quantum Information, California Institute of Technology, Pasadena, CA 91125}
\affiliation{Department of Physics, California Institute of Technology, Pasadena, CA 91125, USA}
\author{Roman~M.~Lutchyn}
\affiliation{Station Q, Microsoft Research, Santa Barbara, CA 93106-6105}
\author{Matthew P.\ A.\ Fisher}
\affiliation{Department of Physics, University of California, Santa Barbara, CA 93106}

\date{\today}
\begin{abstract}
One of the most promising proposals for engineering topological superconductivity and Majorana fermions employs a spin-orbit coupled nanowire subjected to a magnetic field and proximate to an s-wave superconductor.  When only part of the wire's length contacts to the superconductor, the remaining conducting portion serves as a natural lead that can be used to probe these Majorana modes via tunneling.  The enhanced role of interactions in one dimension dictates that this configuration should be viewed as a superconductor-Luttinger liquid junction.  We investigate such junctions between both helical and spinful Luttinger liquids, and topological as well as non-topological superconductors.  We determine the phase diagram for each case and show that universal low-energy transport in these systems is governed by fixed points describing either perfect normal reflection or perfect Andreev reflection.  In addition to capturing (in some instances) the familiar Majorana-mediated `zero-bias anomaly' in a new framework, we show that interactions yield dramatic consequences in certain regimes.  Indeed, we establish that strong repulsion removes this conductance anomaly altogether while strong attraction produces dynamically generated effective Majorana modes even in a junction with a trivial superconductor.  Interactions further lead to striking signatures in the local density of states and the line-shape of the conductance peak at finite voltage, and also are essential for establishing smoking-gun transport signatures of Majorana fermions in spinful Luttinger liquid junctions.
\end{abstract}

\maketitle

\section{Introduction}

Topological phases display many `exact' features that are
remarkably robust to variations in microscopic realization and
imperfections.  One of the most exotic is the possibility of
emergent excitations known as non-Abelian anyons that host
zero-energy `internal' degrees of freedom in an intrinsically
non-local way.  Information encoded in such anyons is then
naturally protected from decoherence and can furthermore be
manipulated using braiding operations, forming the basis of
attractive quantum computing platforms.\cite{TQCreview}  Recently
attention has focused on the so-called Majorana fermion, which
binds to a specific kind of non-Abelian anyon originally predicted
to occur in the $\nu = 5/2$ fractional quantum Hall
state\cite{MooreRead,ReadGreen}.  Among the many current proposals
for generating Majorana
fermions\cite{FuKane, Sau, Alicea} (see [\onlinecite{BeenakkerReview,AliceaReview}] for a review), one attractive class
involves spinless one-dimensional (1D) topological superconductors
whose hallmark is the existence of Majorana zero-modes localized
at the ends of the system.\cite{1DwiresKitaev}  This phase can be
engineered in a variety of settings, including 2D topological
insulator edges\cite{MajoranaQSHedge}, spin-orbit-coupled
nanowires\cite{1DwiresLutchyn,1DwiresOreg}, 3D topological
insulator nanoribbons\cite{MajoranaTInanowires}, and cold atomic
gases\cite{ColdAtomMajoranas}.  Numerous quantum computation
protocols based on manipulating Majoranas in 1D systems have now
been proposed \cite{AliceaBraiding, Hassler,
SauWireNetwork,Flensberg,ClarkeBraiding,
SauBraiding,TopTransmon,JiangKanePreskill,TopologicalQuantumBus,BraidingWithoutTransport,Xue}.

\begin{table*}
\begin{tabular}{|c|c|c|}
\hline
\phantom{blah}& Helical wire & Spinful wire $(g_\sigma=1)$ \\
\hline
\phantom{blah} & \phantom{blah} &   \phantom{blah}\\
Topological SC & $G= \left\{ \begin{array}{rl}
 0, & g < \frac{1}{2} \\
 \frac{2e^2}{h}, & g > \frac{1}{2}.
       \end{array} \right.$ & $G= \left\{ \begin{array}{rl}
 0, & g_\rho < \frac{1}{3} \\
 \frac{2e^2}{h}, & g_\rho > \frac{1}{3}.
       \end{array} \right.$ \\
\phantom{blah} & \phantom{blah} &   \phantom{blah}\\
\hline
\phantom{blah} & \phantom{blah} &   \phantom{blah}\\
Non Topological SC & $G= \left\{ \begin{array}{rl}
 0, & g < 2 \\
 \frac{2e^2}{h}, & g > 2.
       \end{array} \right.$ & $G= \left\{ \begin{array}{rl}
 0, & g_\rho < 1 \\
 \frac{2e^2}{h}, & g_\rho > 1.
       \end{array} \right.$ \\
\phantom{blah} & \phantom{blah} &   \phantom{blah}\\
\phantom{blah} & \phantom{blah} & $ 0 \leq G\leq\frac{4e^2}{h}$,
$g_\rho=1$\\
\phantom{blah} & \phantom{blah} &   \phantom{blah}\\
\hline
\end{tabular}
\caption{The four junction archetypes studied in this paper. The
columns denote the two possibilities for the one dimensional
leads, and the rows differentiate between superconductors with and
without Majorana zero mode bound states. The Luttinger parameters
$g$ and $g_\rho$ describe the interactions in the lead; $g<1$ and
$g>1$ correspond to repulsive and attractive interactions
respectively. Note that for the regime of weak repulsive
interactions expected in many solid state implementations, a zero
bias conductance $G=2e^2/h$ is a robust signature of Majorana zero
mode bound states in the superconductor.}
\end{table*}

The first step towards realizing such applications is of course the conclusive experimental identification of Majorana fermions.  One appealing detection method involves transport.  In particular, several studies predict that tunneling electrons onto a Majorana mode gives rise to a zero-bias conductance anomaly.\cite{ZeroBiasAnomaly0,ZeroBiasAnomaly1,ZeroBiasAnomaly2,ZeroBiasAnomaly3,ZeroBiasAnomaly4,ZeroBiasAnomaly31,ZeroBiasAnomaly5,ZeroBiasAnomaly6,ZeroBiasAnomaly61, ZeroBiasAnomaly7}  Here we revisit this problem from a new perspective based on renormalization group methods similar to those of Refs.~[\onlinecite{KaneFisher1,KaneFisher2,KaneFisher3}], and especially [\onlinecite{Affleck}] \footnote{We would like to thank Ian Affleck for making us aware of Ref. [\onlinecite{Affleck}], where the renormalization group approach was originally applied to the case of a spinful wire.}.  A major virtue of this approach is that it allows one to extract \emph{universal} tunneling signatures of these modes even when strong interactions are present.  Indeed, we will (in some cases) recover in a very general way previous results based on specific model calculations, and also identify new regimes where interactions lead to dramatic and very surprising consequences.  Furthermore, our approach provides an elegant means of addressing the fate of localized Majorana zero-modes when coupled to gapless degrees of freedom.

The theoretical technology developed here is widely applicable to
superconducting Majorana platforms.  We will, however, mainly
focus on spin-orbit-coupled 1D systems such as a semiconducting
nanowire subjected to a magnetic field, in the experimentally
accessible geometry where half of the wire couples to an $s$-wave
superconductor while the other half remains gapless; see, for
example, Fig.\ \ref{TopoSCjunction}(a) and Ref.\
\onlinecite{ZeroBiasAnomaly7,SimonBena}.  By gating one can independently
tune the left and right halves between a `helical' regime---with
only one active channel at low energies---and a `spinful' regime
where two channels play a role (multi-channel regimes are also
accessible~\cite{WimmerMultichannel,Potter_multi, Lutchyn_multi, ZeroBiasAnomaly61}
but will not be considered here).  Crucially, in either limit the
gapless half generically forms a \emph{Luttinger liquid}.  The
nature of the superconducting state in the other half of the wire
depends strongly on the number of channels: in the helical case a
topological phase supporting Majorana zero-modes emerges, while in
the spinful regime a trivial gapped state appears
instead.\cite{1DwiresLutchyn,1DwiresOreg}  This system therefore
admits four natural superconductor/Luttinger liquid junction
archetypes: the gapless part of the wire (which we view as a lead)
can be either helical or spinful and likewise the superconducting
region can be topological or trivial.  We stress that it is
essential to understand the transport properties of both the
topological \emph{and} trivial junctions to establish unambiguous
transport signatures of Majorana modes.  For instance, if the
conductance can behave similarly with or without the presence of a
Majorana then clearly this would be less than a `smoking-gun'
detection scheme.\\


To introduce the basic philosophy underlying our approach it is
useful to first imagine physically cutting the wire such that the
Luttinger liquid and superconducting regions decouple entirely.
Transfer of electrons across the junction formed by the two
subsystems is then prohibited for trivial reasons.  In
renormalization group language, the system's low-energy behavior
here is described by a fixed point theory characterized by a
vanishing conductance.  We refer to this as a `perfect normal
reflection' fixed point since in this case electrons incident on
the superconductor undergo normal reflection with unit probability
at the junction.  Suppose now that the Luttinger liquid and
superconductor `reconnect', and one incorporates \emph{arbitrary}
symmetry-allowed couplings between the two.  Our objective is to
then address questions such as the following: What is the fate of
the perfect normal reflection fixed point in this case?  If it is
unstable, what couplings provide the leading instability and to
which fixed point do they ultimately drive the system?  What are
the properties of such putative fixed points?  And what are the
implications for transport experiments?

Let us now highlight our main results (partially summarized in Table I) for the four cases that we analyze, beginning with the topological superconductor/helical Luttinger liquid junction.  In this case the gapless region is characterized by a Luttinger parameter $g$ where $g = 1$ represents the free-fermion limit while $g<1$ and $g>1$ respectively correspond to repulsive and attractive interactions.  When the wire is cut as described above the topological superconductor supports a single localized Majorana zero-mode at the junction.\footnote{In this work, we use the term topological superconductor in a restricted way to mean a superconductor that supports a single Majorana zero mode at the junction.  } Provided the Luttinger parameter falls in the range $g>1/2$ tunneling electrons onto this mode constitutes a relevant perturbation that destabilizes the perfect normal reflection fixed point.  We demonstrate that the Majorana zero-mode then delocalizes completely into the Luttinger liquid and drives the system to a fixed point describing perfect \emph{Andreev} reflection at the junction.  This perfect Andreev reflection fixed point is characterized by the familiar quantized zero-bias conductance $G = 2e^2/h$ at zero temperature $T=0$ that has been captured by numerous studies in the free-fermion limit.\cite{ZeroBiasAnomaly0,ZeroBiasAnomaly1,ZeroBiasAnomaly2,ZeroBiasAnomaly3,ZeroBiasAnomaly4,ZeroBiasAnomaly5,ZeroBiasAnomaly6,ZeroBiasAnomaly7}
In addition to the universal value of $2e^2/h$ for the conductance at zero bias, the topological superconductor/helical Luttinger liquid junction also exhibits a universal form for the finite bias conductance curve.  The form of this curve can be computed in perturbation theory in certain voltage regimes, allowing one to extract the value of the Luttinger parameter $g$.  For $g<1/2$, however, (which is potentially applicable to carbon nanotube-based Majorana proposals\cite{CNTmajoranas1,CNTmajoranas2,CNTmajoranas3}) coupling to the zero-mode at the junction is irrelevant at the perfect normal reflection fixed point, which is then stable.  Consequently the zero-bias conductance \emph{vanishes} at $T=0$.  This does not, however, imply that the Majorana mode remains localized---we show using scaling that in this limit the probability density associated with the Majorana mode decays into the Luttinger liquid as a power law.

When a helical Luttinger liquid impinges instead on a trivial superconductor, the perfect normal reflection fixed point is stable, resulting in a vanishing zero-bias conductance, for any $g<2$.  Taken together, the above results imply that electron transport from a helical wire (with not-too-strong repulsive interactions) onto a superconductor is unmistakably different in the topological and trivial cases and thus indeed provides an unambiguous way of identifying Majorana zero-modes.  The $g>2$ regime may be realizable in cold atoms experiments\cite{ColdAtomMajoranas} and exhibits fascinating properties even from a purely theoretical perspective.  Here Cooper-pair tunneling from the Luttinger liquid to the superconductor is relevant, generating a flow to the perfect Andreev reflection fixed point with quantized conductance.  We show that this results from the \emph{dynamical} generation of a pair of asymptotically decoupled Majorana modes which mediate the perfect Andreev reflection.  It is remarkable that any trace of Majorana physics appears in a junction with an ordinary superconductor; this would certainly be interesting to explore further in numerical simulations.

Perhaps more than any other case, superconductor/spinful Luttinger liquid junctions highlight the limitations of model-specific microscopic calculations and the utility of our approach based on universal low-energy physics.  When a spinful Luttinger liquid couples to a topological superconductor, the leading perturbation at the perfect normal reflection fixed point again involves electron tunneling onto the Majorana zero-mode at the junction.  Its relevance follows from the charge- and spin-sector Luttinger parameters $g_\rho$ and $g_\sigma$.  With unbroken $SU(2)$ spin symmetry in the Luttinger liquid one has $g_\sigma=1$, and in this case coupling to the Majorana is relevant for all $g_\rho>1/3$.  Because only one of the two spin channels can hybridize with the Majorana mode, the junction flows to a novel fixed point corresponding to perfect Andreev reflection for one species and perfect normal reflection for the other.  This fixed point is in fact robust with respect to some perturbations which break $SU(2)$ in the bulk, {\it i.e.} correspond to $g_\sigma \neq 1$, such as spin orbit coupling and magnetic field.  Thus, in the topological case, a spinful Luttinger liquid junction again exhibits robust $G = 2e^2/h$ conductance quantization down to fairly strong repulsive interactions with $g_\rho=1/3$.

An analysis of the non-topological superconductor/spinful Luttinger liquid case in the free-fermion limit, however, yields a non-universal zero-bias conductance ranging anywhere from $0$ to $4e^2/h$ depending on parameters---potentially making it difficult to differentiate from the signal originating from the Majorana in the topological junction.  This apparent non-universality originates from the fact that the leading perturbations to the perfect normal reflection fixed point in the free fermion case are exactly marginal.  Fortunately we find that arbitrarily weak repulsive interactions are sufficient to stabilize the perfect normal reflection fixed point.  Since weak repulsive interactions are generic in many physical realizations of such Luttinger liquids, the non-universal zero-bias conductance calculated for the free fermion limit will eventually renormalize to zero at sufficiently long length scales and low energies.  We thus conclude that a quantited zero-bias $G=2e^2/h$ conductance indeed serves as a definitive fingerprint of a Majorana mode at the junction, for both spinless \emph{and} spinful Luttinger liquids.

The remainder of the paper analyzes each of the four cases in detail.  We begin in Sec.\ \ref{TopoSC-HelicalLL} with the helical/topological junction, which we first solve in the free fermion limit $g=1$ and then in the general case using bosonization.  In particular, we extract the conductance from a boundary action obtained by integrating out the bulk of the Luttinger liquid, and use duality to gain insight into the nature of the two fixed points.  In Sec.\ \ref{nthl} we apply similar methods to analyze the helical/non-topological junction.  Sections \ref{ntsll} and \ref{tsll} treat the spinful/non-topological and spinful/topological junctions.  In the discussion (Sec. \ref{sec:discussion}) we examine the physical consequences of our work in more detail.  In particular, we show that the line-shape of the finite bias conductance curve contains information about the Luttinger parameter $g$, and that its generic asymptotics ($g \neq 1$) differs from that of the finely tuned free fermion case $g=1$.  Also, Appendices \ref{PhiDeriv} and \ref{PathIntregralDerivation} provide details on the derivation of the bosonized boundary theories, while Appendices \ref{MajoranaSolution} and \ref{interactingMS} solve for the delocalized Majorana mode at the junction in the non-interacting and interacting helical/topological cases, respectively.

\section{Topological superconductor--helical Luttinger liquid junctions}
\label{TopoSC-HelicalLL}

\begin{figure}
\includegraphics[width = 7cm]{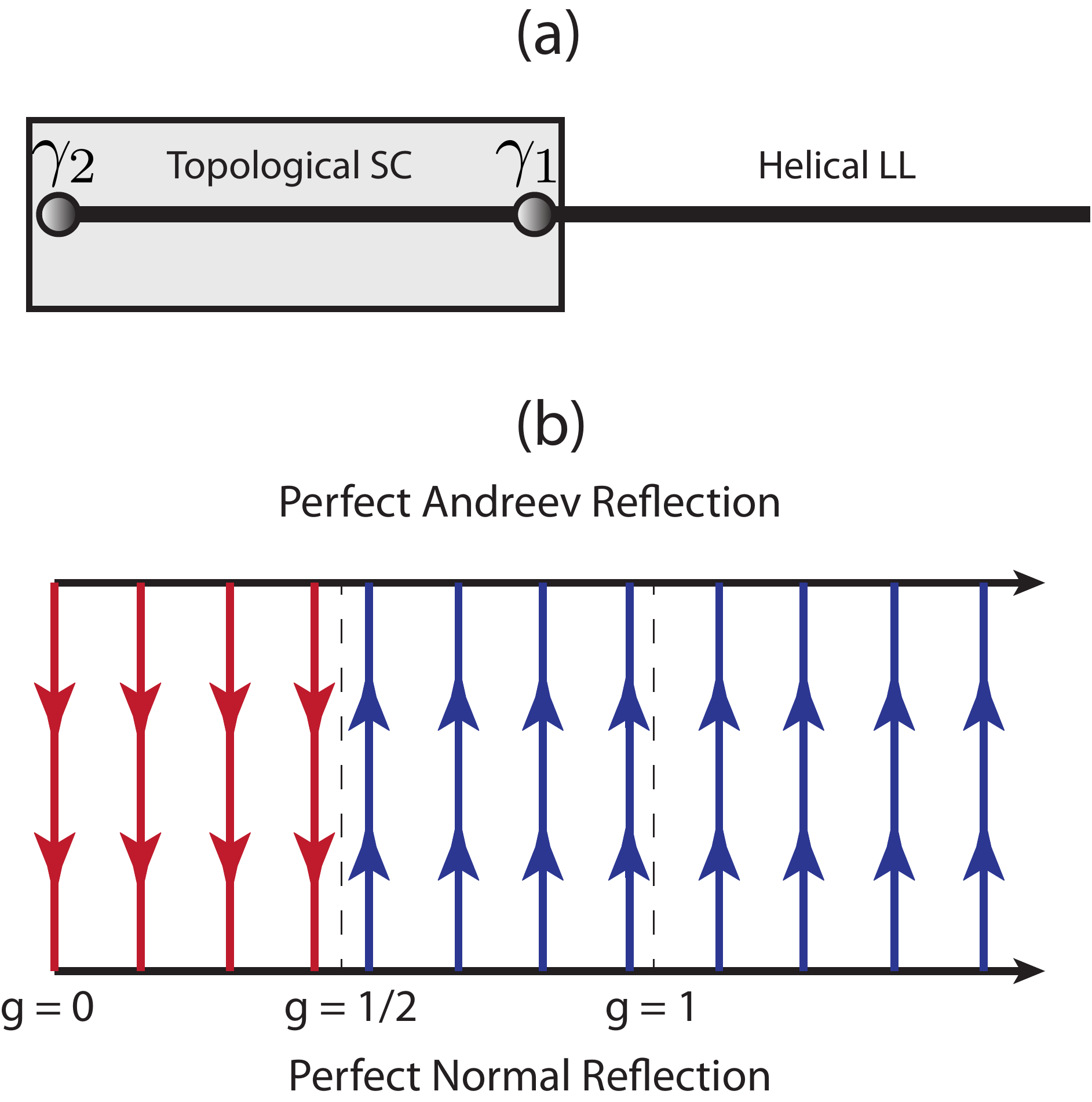}
\caption{(a) Topological superconductor forming a junction with a helical Luttinger liquid.  We assume that the helical Luttinger liquid couples only to the Majorana $\gamma_1$ at the interface, while its partner $\gamma_2$ remains a robust zero-energy mode. (b) Flow diagram for the junction as a function of the interaction parameter $g$ for the Luttinger liquid.  Provided $g > 1/2$, coupling to the Majorana $\gamma_1$ causes the Luttinger liquid to flow onto a fixed point where perfect Andreev reflection occurs at the junction.  Here the zero-bias conductance for the junction is quantized at $2e^2/h$.  For Luttinger liquids with strong repulsive interactions such that $g < 1/2$, however, coupling to $\gamma_1$ is irrelevant.  Here the system flows onto a fixed point where the Luttinger liquid undergoes perfect normal reflection at the junction, leading to a vanishing zero-bias conductance.  }
\label{TopoSCjunction}
\end{figure}

The first junction we will analyze is that formed by a helical Luttinger liquid adjacent to a 1D topological superconductor supporting a single localized Majorana mode at each end as illustrated in Fig.\ \ref{TopoSCjunction}(a).  While ultimately we wish to understand the universal properties of the junction at low energies in the presence of (possibly strong) interactions in the Luttinger liquid, here we will begin by exploring the free fermion case which provides a useful point of reference.

\subsection{Scattering problem for free fermions}
\label{Scattering-Free}

Throughout this section we will assume that the superconductor is fully gapped save for the end-Majorana modes $\gamma_1$ and $\gamma_2$ shown in Fig.\ \ref{TopoSCjunction}(a).  We further assume that the superconductor is sufficiently long that these Majoranas overlap negligibly~\cite{1DwiresKitaev, Meng_splitting}, and that the helical wire couples only to $\gamma_1$.  The only relevant low-energy degrees of freedom are then $\gamma_1$ and those of the gapless helical wire, taken here to be non-interacting.  At low energies it suffices to linearize the kinetic energy for the helical wire and incorporate the effects of the superconductor through local perturbations to the wire Hamiltonian acting at the interface.  (Formally, the latter perturbations can be derived by coupling the wire and superconductor at the junction and then integrating out the gapped superconductor degrees of freedom.)  Taking a semi-infinite wire at $x > 0$, we write the full Hamiltonian as $H = H_0 + \delta H$, where $\delta H$ represents the boundary terms and the kinetic energy $H_0$ reads
\begin{equation}
  H_0 = \int_0^\infty dx\left(-i v_F \psi_R^\dagger \partial_x \psi_R + i v_F \psi_L^\dagger \partial_x \psi_L\right).
  \label{H0}
\end{equation}
Here $v_F$ is the Fermi velocity and $\psi_{R/L}^\dagger$ creates right/left-moving excitations near the Fermi energy.  It is convenient to rewrite $H_0$ in terms of a single fermion field $\psi(x)$ defined over all $x$ as follows:
\begin{eqnarray}
  \psi(x) = \left\{ \begin{array}{rl}
 \psi_R(x), & x > 0 \\
 \psi_L(-x), & x < 0.
       \end{array} \right.
   \label{psi}
\end{eqnarray}
In terms of $\psi(x)$, $H_0$ becomes simply
\begin{equation}
  H_0 = \int_{-\infty}^\infty dx\left(-i v_F \psi^\dagger \partial_x \psi \right).
  \label{H0psi}
\end{equation}

The boundary Hamiltonian $\delta H$ will encode various couplings at the junction, including tunneling between the Majorana mode $\gamma_1$ and the wire, local chemical potential modifications, and Cooper pairing induced locally in the wire by the superconductor.
Thus the full Hamiltonian will take the form of a Bogoliubov-de Gennes equation.  Before turning to a specific form of $\delta H$ it is useful to discuss the problem in some generality.  It is natural to expect that in the low-energy limit the effect of these couplings will be to impose certain boundary conditions on the wire, the precise nature of which will depend on details of the Hamiltonian.  The set of possible boundary conditions corresponds, in renormalization group language, to different fixed points whose stability we would like to understand.  For free fermions this issue can be exhaustively addressed using scattering theory.  Here, all information regarding these putative fixed points as well as their stability is encoded in the S-matrix $S(E)$, which at a given energy $E$ relates the particle and hole states incident on the superconductor with the reflected states.  Interestingly, using arguments similar to Refs.~\onlinecite{FuKaneMajoranaChargeTransport, BeenakkerMajoranaInterferometry} one can show very generally that the S-matrix for an infinite wire can take only one of two forms in the $E\rightarrow 0$ limit, corresponding to fixed points describing either perfect normal reflection, or perfect Andreev reflection at the junction. In the case of a finite wire of length $L$, the S-matrix has been derived in Refs.~[\onlinecite{ZeroBiasAnomaly1, Tewari'08, Nilsson'08}], and one recovers the above answers in the appropriate limit, {\it i.e.} $L=\infty, E\rightarrow 0$.

\subsubsection{General properties of the S-matrix}
\label{SmatrixProperties}

The full Hamiltonian $H$ including boundary couplings is diagonalized with quasiparticle operators carrying energy $E$ of the form
\begin{equation}
  \Gamma_E = \phi_E \gamma_1 + \int_{-\infty}^\infty dx e^{-i \frac{Ex}{v_F}}[P_E(x) \psi(x) + H_E(x) \psi^\dagger(x)],
  \label{Gamma}
\end{equation}
where $\phi_E$ is the component of the wavefunction at the Majorana and $P_E(x), H_E(x)$ respectively determine the particle- and hole-like amplitudes of the wavefunctions.
Since $P_E(-\infty)$ and $H_E(-\infty)$ describe left-moving electrons and holes incident on the superconductor, while $P_E(\infty)$ and $H_E(\infty)$ capture the reflected right-moving states [see Eq.\ (\ref{psi})], the S-matrix is defined by
\begin{eqnarray}
\left[ \begin{array}{c}
  P_E(\infty) \\
  H_E(\infty)
  \end{array}\right]
  &=& S(E) \left[ \begin{array}{c}
  P_E(-\infty) \\
  H_E(-\infty)
  \end{array}\right]
  \label{Smatrix} \\
  &\equiv & \left[ \begin{array}{cc}
  S_{PP}(E) & S_{PH}(E) \\
  S_{HP}(E) & S_{HH}(E)
  \end{array}\right] \left[ \begin{array}{c}
  P_E(-\infty) \\
  H_E(-\infty)
  \end{array}\right].
  \nonumber
\end{eqnarray}
To restrict the form of $S(E)$, we first note that $\Gamma_E = \Gamma_{-E}^\dagger$, which follows from particle-hole symmetry exhibited by any Bogoliubov-de Gennes equation, combined with the fact that the helical wire has only a single fermionic species.  This relation connects the particle and hole wavefunction amplitudes via $P_{-E}^*(x) = H_E(x)$, which in turn implies that the S-matrix must obey $S(E) = \sigma^x S^*(-E) \sigma^x$, where $\sigma^x$ is a Pauli matrix.  Thus it suffices to determine (say) $S_{PP}(E)$ and $S_{PH}(E)$, since the other matrix elements follow from
\begin{eqnarray}
  S_{HH}(E) &=& S_{PP}^*(-E)
  \nonumber \\
  S_{HP}(E) &=& S_{PH}^*(-E).
  \label{SmatrixElementsRelation}
\end{eqnarray}

At zero energy, Eqs.\ (\ref{SmatrixElementsRelation}) allow one to restrict the S-matrix further still.  Imposing $S_{PP}(0) = S_{HH}^*(0)$ and $S_{PH}(0) = S_{HP}^*(0)$ along with unitarity of the S-matrix as required by current conservation, one finds that $S(E = 0)$ can take \emph{only} two possible forms:
\begin{equation}
S(0) = \left( \begin{array}{cc}
  e^{i\alpha} & 0 \\
  0 &  e^{-i\alpha}
  \end{array}\right), ~(\text{perfect normal reflection})
  \label{ReflectingBCs}
\end{equation}
or
\begin{equation}
S(0) = \left( \begin{array}{cc}
  0 & e^{i\beta} \\
  e^{-i\beta} & 0
  \end{array}\right), ~(\text{perfect Andreev reflection}),
  \label{AndreevBCs}
\end{equation}
for some phases $\alpha$ and $\beta$.
The diagonal form in Eq.\ (\ref{ReflectingBCs}) corresponds to the situation where particles incident on the superconductor undergo perfect normal reflection at the interface---an electron reflects as an electron with unit probability and similarly for holes.  Conversely, the purely off-diagonal form in Eq.\ (\ref{AndreevBCs}) describes perfect Andreev reflection at the junction; here electrons scatter perfectly into holes and vice versa, transmitting a Cooper pair into the superconductor in the process.  In renormalization group terms, these limits correspond to two different fixed points at which the topological superconductor imposes either perfect normal reflecting or perfect Andreev reflecting boundary conditions on the helical wire.  Given our assumptions fixed points with intermediate boundary conditions are not possible (at least for free fermions).

Following Ref.\ \onlinecite{ZeroBiasAnomaly3}, these two possible fixed points can be distinguished by the conductance across the junction.  At a bias voltage $V$, the current transmitted into the superconductor is given by
\begin{equation}
  I = \frac{2e}{h} \int_0^{eV}dE |S_{PH}(E)|^2,
\end{equation}
where $|S_{PH}(E)|^2$ is the probability that an incident electron at energy $E$ Andreev reflects into a hole at the junction, transmitting charge $2e$ into the superconductor.  The differential conductance $G = \frac{dI}{dV}$ at $T=0$ is then
\begin{equation} \label{difG}
  G = \frac{2e^2}{h}|S_{PH}(eV)|^2,
\end{equation}
which in the zero-bias limit becomes
\begin{equation}
  G(V\rightarrow 0) = \left\{ \begin{array}{c}
 0 ~~~~\text{(perfect normal reflection)} \\
 \frac{2e^2}{h} ~~~~  \text{(perfect Andreev reflection)}.
       \end{array} \right.
\end{equation}
While it is intuitively clear that perfect normal reflection ought to give rise to a vanishing zero-bias conductance, it is interesting that in the Andreev limit one necessarily obtains conductance quantization.  We will return to this issue below when we obtain the S-matrix for a specific form of $\delta H$.

\subsubsection{Accessing the perfect normal and Andreev reflection fixed points with free fermions}
\label{FixedPointsFreeFermions}

We would like to now understand the conditions required for our non-interacting helical wire to flow onto each of the two possible fixed points identified above.  With this objective in mind we will now consider the following boundary Hamiltonian:
\begin{eqnarray}
  \delta H &=& \int _{-\infty}^\infty dx \bigg{[}\frac{t}{\sqrt{2}} \gamma_1(\psi^\dagger-\psi) + 2u\psi^\dagger \psi
  \nonumber \\
  &+& \left(i \Delta\psi \partial_x \psi + h.c.\right)\bigg{]}\delta(x).
  \label{deltaH}
\end{eqnarray}
Here, $t$ allows electron tunneling between the Majorana $\gamma_1$ and the helical wire at the junction, $u$ is a local potential that favors normal reflection, and the $\Delta$ term (which must involve a derivative by Fermi statistics) encodes processes wherein a Cooper pair hops between the helical wire and the superconductor.  Additional couplings are in principle present but necessarily carry higher derivatives than those already displayed and can thus be safely neglected at low energies.
Given this form of $\delta H$, a straightforward solution of the S-matrix yields,
\begin{eqnarray}
  S_{PP} &=& -\frac{\tilde E [(\tilde u+i)^2 + \tilde \Delta(\tilde t^2  +
    \tilde E^2 \tilde \Delta)]}{
 i \tilde t^2 + \tilde E[1 +
     \tilde u^2 + \tilde \Delta (\tilde t^2 + \tilde E^2 \tilde \Delta)]}
     \nonumber \\
   S_{PH} &=& \frac{i (\tilde t^2 + 2 \tilde E^2 \tilde \Delta)}{ i \tilde t^2 + \tilde E[1 +
     \tilde u^2 + \tilde \Delta (\tilde t^2 + \tilde E^2 \tilde \Delta)]},
     \label{SmatrixAndreev}
\end{eqnarray}
where the tildes denote quantities expressed in units of $v_F$, \emph{e.g.}, $\tilde u = u/v_F$.  
[Recall that the other two matrix elements follow from Eqs.\ (\ref{SmatrixElementsRelation}).]
For generic values of the couplings one clearly sees that the S-matrix becomes purely off-diagonal in the $E\rightarrow 0$ limit, corresponding to the onset of perfect Andreev reflection at the boundary and a quantized zero-bias conductance of $2e^2/h$ for the junction.  Avoiding this outcome requires fine-tuning, indicating that the Andreev fixed point is stable in the non-interacting case.

Closer inspection of Eqs.\ (\ref{SmatrixAndreev}) reveals that the flow towards perfect Andreev boundary conditions originates exclusively from the coupling to the Majorana $\gamma_1$ located at the junction.  Indeed, in the $E\rightarrow 0 $ limit the S-matrix becomes purely diagonal upon fine-tuning $t = 0$, so that without coupling to the Majorana the system flows instead onto the (unstable) perfect normal reflection fixed point.  (One might naively expect that the pairing term $\Delta$ alone would be sufficient to drive the system to the Andreev fixed point, but this is not the case.  In the helical wire, Pauli exclusion, which is responsible for the derivative in this term, renders this an irrelevant perturbation at the perfect normal reflection fixed point.  This can be understood by noting that when $E \rightarrow 0$, $\Delta$ drops out entirely from the S-matrix.)  It follows that here the quantized $2e^2/h$ conductance for the junction at the Andreev fixed point reflects the familiar zero-bias anomaly associated with tunneling onto a Majorana mode\cite{ZeroBiasAnomaly0,ZeroBiasAnomaly1,ZeroBiasAnomaly2,ZeroBiasAnomaly3,ZeroBiasAnomaly4,ZeroBiasAnomaly5,ZeroBiasAnomaly6,ZeroBiasAnomaly7}.  The scattering approach adopted here in fact follows closely the treatment of Ref.\ \onlinecite{ZeroBiasAnomaly3} who showed that resonantly coupling a non-interacting system to a Majorana mode generically induces perfect Andreev reflection as recovered here.  One can intuitively understand this by observing that for any non-zero $t$, the Majorana zero-mode originally described by $\gamma_1$ gets absorbed into the helical wire, where it becomes a delocalized plane-wave state (see Appendix \ref{MajoranaSolution}).  The Majorana character of this plane-wave must be preserved, however, which in turn guarantees perfect Andreev reflection.

Next we explore the stability of the fixed points captured here for free fermions when interactions are present.  Interestingly, we will show that strong repulsive interactions modify the physics of the junction qualitatively.  We will first treat the perfect normal reflection fixed point and then turn to the Andreev fixed point.

\subsection{Stability of the perfect normal reflection fixed point with interactions}
\label{NormalFixedPointStability}

In the non-interacting limit accessing the perfect normal reflection fixed point required fine-tuning to zero the coupling between the helical wire and the Majorana $\gamma_1$ at the junction.  Guided by this case, we will initially neglect the presence of $\gamma_1$ and derive a fixed-point action describing perfect normal reflection for the Luttinger liquid.  This is conveniently achieved using bosonization, where the right/left-moving fermionic modes in the wire are expressed in terms of dual bosonic fields $\phi, \theta$ via
\begin{eqnarray}
  \psi_R &\sim& e^{i(\phi + \theta)}
  \\
  \psi_L &\sim& e^{i(\phi - \theta)}.
\end{eqnarray}
Physically, $\theta$ relates to the fermion density $n$ in the
Luttinger liquid according to $n = \partial_x\theta/\pi$,
while $\phi$ and $n$ are canonically conjugate variables.  It will
be useful below to observe that since one can write
$\theta(x)-\theta(0) = -\pi \int_0^x dx'n(x')$, the fermion parity
in the helical Luttinger liquid is given by $P_{LL} =
\cos[\theta(x = L)-\theta(x = 0)]$, where $x=L$ corresponds to the
right endpoint of the wire.

To ensure perfect normal reflection at the junction the fermionic fields are constrained to satisfy
\begin{equation}
  \psi_R(x = 0) = e^{i\alpha}\psi_L(x = 0)
  \label{NormalBC}
\end{equation}
for some unimportant phase $\alpha$ which we will simply set to zero.  This in turn implies pinning of the field $\theta$ at the junction:
\begin{equation}
  \theta(x = 0) = 0~{\rm mod}~\pi.
  \label{theta_pinning}
\end{equation}
Similar constraints of course apply to the right end of the wire at $x = L$, so for concreteness we will henceforth set $\theta(x = L) = 0$ (except in Appendix \ref{interactingMS}, where a different convention is specified) .  The fermion parity in the Luttinger liquid then reduces to
\begin{equation}
  P_{LL} = \cos[\theta(x = 0)].
  \label{PLL}
\end{equation}
Thus the two pinning values in Eq.\ (\ref{theta_pinning}) correspond to the cases where the Luttinger liquid accommodates an even and odd number of electrons.

Let us now arbitrarily select a particular pinning value for $\theta(x=0)$.  One can obtain an effective action for the remaining fluctuating phase field at the boundary, $\phi(x = 0)$,  in the following manner.  First, the kinetic energy in Eq.\ (\ref{H0}) supplemented by interactions in the helical wire bosonizes to
\begin{eqnarray}
  H_0 = \int_0^\infty dx\frac{v_F}{2\pi}[g(\partial_x \phi)^2 + g^{-1}(\partial_x \theta)^2],
  \label{H0bosonized}
\end{eqnarray}
where $v_F$ is the Fermi velocity and $g$ is the Luttinger parameter specifying the interaction strength (again, $g = 1$ is the free-fermion limit while $g < 1$ and $g > 1$ respectively correspond to repulsive and attractive interactions).  Obtaining the Euclidean action corresponding to Eq.\ (\ref{H0bosonized}) and integrating out all fields away from $x = 0$ (see Appendix \ref{PhiDeriv} for details) then leads to the following action,
\begin{equation}
  S_{\rm normal} = \frac{g}{2\pi} \int \frac{d\omega}{2\pi} |\omega||\Phi|^2,
  \label{NormalFixedPointAction}
\end{equation}
where $\Phi \equiv \phi(x = 0)$.  This action describes the perfect normal reflection fixed point, whose stability we can now assess.

The most relevant perturbation to $S_{\rm normal}$ originates from coupling to the neglected Majorana mode $\gamma_1$,
\begin{eqnarray}
  \delta S_{t} &=& \frac{t}{\sqrt{2}}\int d\tau \gamma_1[\psi_R(x = 0)^\dagger - \psi_R(x = 0)]
  \label{deltaSt}
\end{eqnarray}
which promotes Andreev processes at the junction.
[The use of $\psi_R$ as opposed to $\psi_L$ here is immaterial because of the boundary condition of Eq.\ (\ref{NormalBC}).]  Bosonizing $\delta S_t$ requires some care, as the usual procedure of naively replacing $\psi_R \sim e^{i\Phi}$ in Eq.\ (\ref{deltaSt}) leaves one with a non-Hermitian operator (among other technical problems stemming from the Majorana operator).  Indeed, since Eq.\ (\ref{deltaSt}) contains $\gamma_1$ our bosonization must include the topological superconductor as well in order to obtain consistent results.  We will now show how this can be done by considering a lattice model that includes the relevant low-energy operators for both subsystems.

Because the only low-lying degrees of freedom in the superconductor are $\gamma_1$ and $\gamma_2$, we can distill this part down to a single fermionic lattice site taken to lie at position $0$:
\begin{align} \gamma_1 &= c_0^\dag + c_0 \nonumber \\ \gamma_2 &= i(c_0^\dag - c_0). \end{align}
Even though we formally model $\gamma_1$ and $\gamma_2$ as deriving from the same site, physically these operators are spatially well-separated.  For our purposes the only consequence of this is that the Luttinger liquid, which we now define on a lattice indexed by sites $j>0$, couples only to the linear combination $c_{0}^\dagger + c_{0}$ on site $0$.  We therefore consider the following tunneling Hamiltonian that hybridizes this site and the Luttinger Liquid,
\begin{equation}
\delta H_t = t\gamma_1(c_1^\dagger - c_1) = t(c_{0}^\dagger + c_{0})(c_1^\dagger-c_1).
\label{Ht}
\end{equation}
The full Hamiltonian is then \begin{equation} H = H_{LL} + H_t, \end{equation} with the Luttinger Liquid Hamiltonian \begin{equation}
  H_{LL} = -J \sum_{j > 0}(c_j^\dagger c_{j+1} + h.c.) + H_{\rm int},
\end{equation}
where $H_{\rm int}$ encodes density-density interactions.

Next we implement a Jordan-Wigner transformation to write the Hamiltonian in terms of hard-core bosons $b_j$ via
\begin{equation}
  c_j = \exp\left(i\pi\sum_{j' < j}n_{j'}\right)b_j,
  \label{JordanWigner}
\end{equation}
with $n_j = b_j^\dagger b_j = c_j^\dagger c_j$.  Furthermore, we rewrite the bosons at site $0$ in terms of Pauli spin matrices via $b_0 = (\sigma^x + i \sigma^y)/2$, and those on the remaining sites $j>0$ in terms of a phase field $\phi_j$: $b_j \sim ie^{i \phi_j}$.  Using $\exp\left(i\pi n_0\right) = \sigma^z$ the tunneling term $H_t$ in Eq.\ (\ref{Ht}) then becomes
\begin{equation}
\delta H_t = t \sigma^x \cos \Phi,
\label{Ht_final}
\end{equation}
where $\Phi \equiv \phi_1$.  The Euclidean action corresponding to Eq.\ (\ref{Ht_final}) is
\begin{equation}
  \delta S_t = 2t \int d\tau \sigma^x \cos\Phi,
\label{dstfinal}
\end{equation}
which upon trivially rescaling $t$ is the correctly bosonized form of the Majorana tunneling term in Eq.\ (\ref{deltaSt}).

For concreteness, it is useful to relate the wavefunctions for the superconductor in the fermionic and `spin' languages.  When $t = 0$ the superconductor admits two degenerate ground states with well-defined but opposite fermion parity, $|0\rangle$ and $|1\rangle$, due to the Majorana zero-modes $\gamma_1$ and $\gamma_2$.  These states are connected by the Majorana operators: $\gamma_i|0\rangle \propto |1\rangle$.  In the spin language $|0\rangle$ and $|1\rangle$ are eigenstates of $\sigma^z$, but the natural pair of degenerate ground states is formed by eigenstates $|+\rangle$ and $|-\rangle$ of $\sigma^x$.  The distinction between the two can be sharpened by considering a more realistic model for the superconductor, consisting of a Kitaev model \cite{1DwiresKitaev} with many sites intervening between $\gamma_1$ and $\gamma_2$.  The ground states of given fermion parity, natural in the original fermionic representation, are then Schrodinger cat states for the spins, i.e. linear combinations of the two phase eigenstates, in which all spins point along the plus or minus $x$ direction. \footnote{See [\onlinecite{LSM}] for exact solutions of the spin models corresponding to the Kitaev chain.}
 (Incidentally this is essentially why a 1D spinless $p$-wave superconductor is a widely sought topological phase of matter while the Ising spin chain is not, despite the fact that these models are superficially related.)

We are now in position to analyze the stability of the perfect normal reflection fixed point.  
The scaling dimension of $\cos\Phi$ is $1/(2g)$ while that of $\sigma^x$ is zero at this fixed point, so under renormalization $t$ flows according to
\begin{equation}
  \frac{d t}{d \ell} = [1-(2g)^{-1}]t.
  \label{tflow}
\end{equation}
Equation (\ref{tflow}) determines the renormalized coupling strength at a length scale $l$ in terms of $\ell = \ln(l/l_0)$, with $l_0$ a microscopic length of order the Fermi wavelength.  Remarkably, for helical Luttinger liquids with $g < 1/2$ tunneling onto $\gamma_1$ thus constitutes an \emph{irrelevant} perturbation; perfect normal reflection is then stable despite the presence of a zero-energy Majorana mode to which the system can couple.  We will explore the physical consequences of this result in the discussion.  This coupling is relevant, however, when $g > 1/2$, indicating instability of the perfect normal reflection fixed point (consistent with our scattering analysis for free fermions with $g = 1$).  Since this perturbation promotes Andreev reflection at the junction it is natural to expect that at low energies normal reflection then becomes entirely suppressed in favor of Andreev processes, just as we found for free fermions.  We establish in the next subsection that this is indeed the case by examining the stability of the perfect Andreev reflection fixed point when interactions are present.

\subsection{Stability of the perfect Andreev reflection fixed point with interactions}
\label{AndreevFixedPointStability}

One illuminating method for extracting the fixed point action describing perfect Andreev reflection at the junction is to apply a duality transformation to the bosonized action $S_{\rm normal} + \delta S_{t}$.  In particular, this allows us to extract the most relevant operator around this fixed point and analyze its stability.  Our starting point is the partition function for the perfect normal reflection fixed point expressed as a path integral; this is carefully derived in Appendix \ref{PathIntregralDerivation} and reads
\begin{eqnarray}
  Z = \int \mathcal{D} \Phi \sum_{\sigma^x = \pm 1}e^{-S_{\rm normal}}e^{-2t \int d\tau \sigma^x\cos\Phi}
  \label{Z}
\end{eqnarray}
An important point here is that there is only a \emph{single} sum over $\sigma^x = \pm 1$ (rather than one at each imaginary time slice).  This arises from the fact that $\sigma^x$ is a conserved quantity in the Hamiltonian and therefore has no imaginary time dynamics.  Upon writing the cosine term in the Villain approximation the partition function then becomes
\begin{eqnarray}
  Z &\approx& \int \mathcal{D}\Phi \sum_{\sigma^x = \pm 1}\sum_{\{n(\tau)\}\in \mathbb{Z}}e^{-S_{\rm normal}}
  \nonumber \\
  &\times& e^{-t \int d\tau [\Phi + (1+\sigma^x)\pi/2-2\pi n]^2}.
  \label{Zvillain}
\end{eqnarray}
In Villainized form the $t$ term can be decoupled with a Hubbard-Stratonovich field $\rho$, yielding
\begin{eqnarray}
  Z &=& \int \mathcal{D}\Phi \mathcal{D}\rho \sum_{\sigma^x = \pm 1}\sum_{\{n(\tau)\}\in \mathbb{Z}}e^{-S_{\rm normal}}
  \nonumber \\
  &\times& e^{-\int d\tau \left\{\frac{\rho^2}{t} +2i\rho[\Phi + (1+\sigma^x)\pi/2-2\pi n]\right\}}.
  \label{Zrho}
\end{eqnarray}
At this point it is convenient to introduce the variable $\Theta$ which is canonically conjugate to $\Phi/\pi$ by writing
\begin{equation}
  \rho = \frac{\partial_\tau \Theta}{2\pi}.
\end{equation}
Performing the sum over integers $n(\tau)$ in Eq.\ (\ref{Zrho}) restricts the field $\Theta$ to elements of $\pi \mathbb{Z}$, while the $\sigma^x$ sum simply imposes periodic boundary conditions on $\Theta$ along the imaginary time direction [\emph{i.e.}, $\Theta(\tau = \beta) = \Theta(\tau = 0)$ mod $2\pi$].  As usual the former restriction is difficult to handle so we impose the constraint `softly' by adding a potential to the action which energetically favors integer values for $\Theta/\pi$; the partition function then reads
\begin{align}
  Z &\approx  \int \mathcal{D}\Phi \mathcal{D}\Theta e^{-S_{\rm normal}}e^{-\int d\tau \left[\frac{(\partial_\tau \Theta)^2}{(2\pi)^2t} +\frac{i}{\pi}\Phi \partial_\tau \Theta - v \cos(2\Theta)\right]}.
\end{align}
One can then integrate out $\Phi$ to obtain the desired dual theory for $\Theta$:
\begin{eqnarray}
  Z &=& \int \mathcal{D}\Theta e^{-S_{\rm dual}}
  \\
  S_{\rm dual} &=& \int \frac{d\omega}{2\pi}\frac{|\omega|}{2\pi g}|\Theta|^2-v \int d\tau \cos(2\Theta),
  \label{Sdual}
\end{eqnarray}
where we dropped a term proportional to $(\partial_\tau\Theta)^2$ since it is irrelevant compared to the first term in the action above.

Since $\Theta$ and $\Phi/\pi$ are conjugate variables, the operator $e^{2i\Theta}$ shifts $\Phi$ by $2\pi$.  Thus the $v$ term above represents an instanton operator which tunnels between adjacent minima of the $\sigma^x\cos\Phi$ potential in Eq.\ (\ref{Z}).  [One might naively expect instanton operators that simultaneously change $\sigma^x \rightarrow -\sigma^x$ and $\Phi \rightarrow \Phi + \pi$ to be important, but these are forbidden since $\sigma^x$ is a non-fluctuating classical degree of freedom.]  Tunneling events imposed by $\cos(2\Theta)$ are qualitatively unimportant at `large' $t$ but must be retained otherwise.  When $t$ is relevant and the perfect normal reflection fixed point is unstable, the fixed point described by the dual theory with $v = 0$ should therefore be stable and vice versa.  This strongly suggests that $S_{\rm dual}$ with $v = 0$ describes the perfect Andreev reflection fixed point.  We will now confirm this by rederiving $S_{\rm dual}$ beginning from the fermionic theory.

To access the perfect Andreev reflection fixed point, tunneling onto the Majorana $\gamma_1$ (which as we saw in Sec.\ \ref{FixedPointsFreeFermions} underlies the flow to this fixed point) must be incorporated non-perturbatively.  The coupling to $\gamma_1$ constrains the fermionic fields at the junction such that at low energies
\begin{equation}
  \psi_R^\dagger(x = 0) = e^{i\beta}\psi_L(x = 0),
  \label{AndreevBC}
\end{equation}
which upon setting $\beta = 0$ for simplicity pins the bosonized phase field $\phi$ to
\begin{equation}
  \Phi = \phi(x = 0) = 0~{\rm or}~\pi.
  \label{twochoices}
\end{equation}
(See Appendix \ref{MajoranaSolution} for an explicit solution that derives these boundary conditions in the non-interacting limit.)  The appearance of two possible pinning values for $\Phi$ can be understood from our analysis of the perfect normal reflection fixed point above.  There we showed that, by bosonizing the low-energy degrees of freedom for the Luttinger liquid \emph{and} the topological superconductor, the coupling to $\gamma_1$ can be written as $\delta H_t \propto t \sigma^x \cos\Phi$, where $\sigma^x$ swaps between the two opposite-parity degenerate ground states $|0\rangle$ and $|1\rangle$ for the superconductor.  Note that $\delta H_t$ commutes with the total fermion parity operator
\be P_{\rm{tot}} = \sigma^z \cos \theta(x=0), \ee
with $\sigma^z = i \gamma_1\gamma_2$, as expected since the hopping preserves the global parity of the system.
For $g > 1/2$ we showed that tunneling onto $\gamma_1$ constitutes a relevant perturbation at the perfect normal reflection fixed point.  The system then flows at low energies onto a fixed point at which $\sigma^x \cos \Phi$ is pinned to $-1$ (assuming $t>0$).  This allows two possibilities:
\bea \label{twosectors}
  \sigma^x &=&+1, \qquad \cos \Phi =-1 \nonumber \\
  \sigma^x&=&-1, \qquad \cos\Phi=+1,
\eea
corresponding to the two pinning values in Eq.\ (\ref{twochoices}).

It is instructive to examine the ground states corresponding to the two sectors identified above.  At the Andreev fixed point these may be written as
\bea \label{decomp1}
  |\uparrow\rangle&=&|\sigma^x=1\rangle\otimes |\Phi=\pi\rangle \nonumber\\
  |\downarrow\rangle&=& |\sigma^x=-1\rangle \otimes |\Phi=0\rangle,
\eea
where $|\sigma^x = \pm 1\rangle = |0\rangle \pm |1\rangle$, and $|\Phi=0,\pi\rangle$ are the ground states of the bosonized Hamiltonian in Eq.\ (\ref{H0bosonized}) with boundary conditions $\phi(x=0)=0,\pi$. Appropriate linear combinations of the states in Eqs.\ (\ref{decomp1}) yield ground states with well-defined parity $P_{\rm tot}$,
 \bea \label{decomp2}
  |-\rangle&=&  |\uparrow\rangle - |\downarrow \rangle \nonumber\\
  |+\rangle&=&  |\uparrow\rangle + |\downarrow\rangle,
\eea
satisfying $P_{\rm{tot}} |\pm\rangle = \pm |\pm \rangle$.
While the overall parity $P_{\rm{tot}}$ is a good quantum number, the individual parities of the superconductor and the Luttinger liquid are entangled in the two ground states $|\pm\rangle$.  Explicitly, they can be written as
\bea \label{decomp}
  |-\rangle &=& |0\rangle\otimes |P_{\rm LL}=-1\rangle+ |1\rangle \otimes |P_{\rm LL}=1\rangle \nonumber\\
 |+\rangle &=&  |0\rangle\otimes |P_{\rm LL}=1\rangle + |1\rangle \otimes |P_{\rm LL}=-1\rangle,
\eea
Here  $|P_{\rm LL}=\pm 1\rangle = |\Phi=\pi\rangle \pm |\Phi=0\rangle$ have well defined parity in the Luttinger liquid.  We note that the decomposition (\ref{decomp}) is only exact in the limit of a semi-infinite Luttinger liquid and when the Andreev fixed point has been reached.

In fermionic langauge, the leading perturbation away from the Andreev fixed point is the potential term
\begin{equation}
  \delta S_{u} = 2u \int d\tau \left[\psi_R^\dagger(x = 0) \psi_L(x = 0) + {\rm h.c.}\right]
  \label{deltaSnormal}
\end{equation}
which promotes normal reflection at the junction.  Because of the Andreev boundary condition $\psi_R^\dag (x=0)= \psi_L(x=0)$ the definition of $\delta S_u$ above is somewhat subtle.  When regularized at length scale $\epsilon$ (for example via point-splitting) the operator $\psi_R^\dag(x=0) \psi_L(x=0)$ scales to $0$ linearly with $\epsilon$, so a divergent factor of $1/\epsilon$ must be absorbed in $u$ to obtain a finite result.  The bosonized form of this finite coupling is
\begin{equation}
  \delta S_{u} \sim 4u \int d\tau \cos(2\Theta),
  \label{Su}
\end{equation}
which is just the instanton operator in the dual action of Eq.\ (\ref{Sdual}) obtained earlier by complementary means.

Since $\cos(2\Theta)$ has scaling dimension $2g$, to leading order the coupling $u$ flows according to
\begin{equation}
  \frac{d u}{d \ell} = (1-2g)u.
  \label{uflow}
\end{equation}
Provided $g>1/2$ this perturbation is therefore irrelevant and the Andreev fixed point is stable.  Note that this regime includes the non-interacting limit, $g = 1$, consistent with our scattering analysis above.  For $g=1/2$ the perturbation turns out to be exactly marginal: there exists a line of RG fixed points which interpolate between normal and Andreev reflection.

For a strongly repulsive wire with $g < 1/2$, however, the potential $u$ is relevant.  The Andreev fixed point is then unstable towards the perfect normal reflection fixed point analyzed previously (which we found is stable in this range of $g$).  With $g < 1/2$ the backscattering term in Eq.\ (\ref{Su}) pins $\cos(2\Theta)$ to $-1$ at low energies (assuming $u < 0$ for concreteness), yielding the two possible pinning values of $\theta(x = 0)$ identified at the perfect normal reflection fixed point in Eq.\ (\ref{theta_pinning}).  Because of the boundary condition of fixed $\theta(x=L)$ these two pinning values, corresponding to different fermionic parities on the wire, are not degenerate, and in fact are split by an energy of order $1/L$.

It is interesting to ask about the fate of the Majorana zero-mode at the junction in this strongly repulsive regime.  Although the tunneling $t$ between $\gamma_1$ and the Luttinger liquid is irrelevant for $g < 1/2$, this term nevertheless has a quantitative effect on the zero-mode operator since any finite $t$ makes the commutator $[H,\gamma_1]$ non-zero.  Let $\gamma_1^{\rm new}$ be the operator for the Majorana mode that arises when $t \neq 0$.  In principle this operator can be determined by requiring that $[H,\gamma_1^{\rm new}] = 0$.  We will alternatively deduce the asymptotic form of the probability density $P(x)$ associated with $\gamma_1^{\rm new}$ using scaling.  [It is tempting to view $P(x)$ as deriving from the Majorana wavefunction corresponding to $\gamma_1^{\rm new}$---which is certainly legitimate in the free-fermion limit but rather subtle in the interacting case.  In the regime of interest here with $g<1/2$, $\gamma_1^{\rm new}$ does not generally admit a single-body expansion in terms of microscopic fermion operators.  Reference \onlinecite{Miles} discusses this issue and demonstrates that even for a strongly interacting system a Majorana `wavefunction' yielding a probability $P(x)$ can be extracted from matrix elements of a fermion operator at position $x$ with respect to opposite-parity ground states.]

Dimensional analysis together with Eq.\ (\ref{tflow}) lead to the following scaling ansatz for $P(x)$,
\begin{equation}
  P(x) = \frac{1}{L}\mathcal{P}\left(x/L; t L^{1-(2g)^{-1}}\right),
\end{equation}
where $L$ is the length of the Luttinger liquid.  Note that $\mathcal{P}$ is a symmetric function of $t$ that vanishes when $t = 0$.  For $g<1/2$ it suffices to treat $t$ perturbatively since this coupling is irrelevant; at lowest nontrivial order one obtains
\begin{equation}
  P(x) \sim \frac{1}{L}\left[t L^{1-(2g)^{-1}}\right]^2 \tilde{\mathcal{P}}(x/L).
\end{equation}
In the perturbative regime we expect the Majorana wavefunction to be normalizable in the $L\rightarrow \infty$ limit (as opposed to plane-wave-like as it is when $t$ is relevant).  Under this assumption the probability density $P(x)$ must be independent of $L$ in the thermodynamic limit which requires the asymptotic form
\begin{equation}
  \tilde{\mathcal{P}}(x/L) \sim \left(\frac{x}{L}\right)^{1-1/g}.
  \label{Ptildeprime}
\end{equation}
As an important self-consistency check, we note that for $g = 1/2$ (where $t$ is marginal) Eq.\ (\ref{Ptildeprime}) yields $P(x) \sim 1/x$ so that the Majorana wavefunction is only quasi-normalizable.  For any $g$ below 1/2, however, one obtains a normalizable probability distribution consistent with our ansatz.  Thus we conclude that in the $g < 1/2$ regime the irrelevant tunneling between $\gamma_1$ and the Luttinger liquid results in a Majorana zero-mode that bleeds into the wire but remains \emph{power-law localized} to the interface.  It would be very interesting to test these predictions in DMRG simulations by adapting the techniques of Ref.\ \onlinecite{Miles} to `see' the Majorana wavefunction in this geometry in a numerical experiment.

We also note as an aside that the critical value of $g = 1/2$ coincides with the critical Luttinger parameter below which pairing induced by proximity in the \emph{bulk} of a helical Luttinger liquid is an irrelevant perturbation\cite{MajoranaInteractions}.  Thus helical wires with $g < 1/2$ not only resist Majorana modes imposed externally by a topological superconductor as in the junction studied here, but also reject pairing that would make the wire itself topologically superconducting.

\section{Non-topological superconductor--helical Luttinger liquid junctions}
\label{nthl}

\begin{figure}
\includegraphics[width = 7cm]{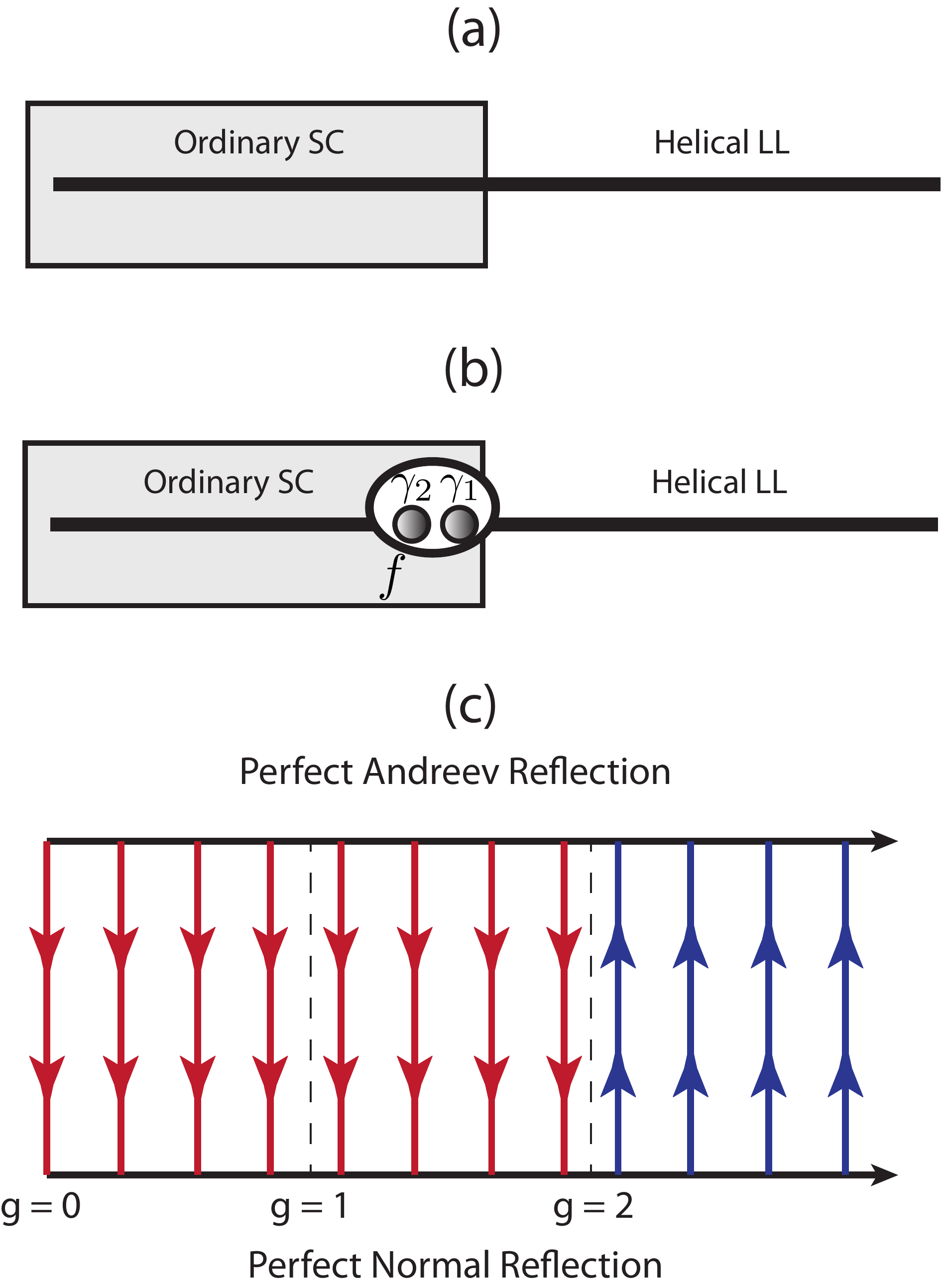}
\caption{(a) Ordinary superconductor forming a junction with a helical Luttinger liquid.  If the superconductor is fully gapped throughout, then in the non-interacting limit the helical wire necessarily flows to a fixed point at which perfect normal reflection occurs at the junction.  Accessing the perfect Andreev reflection fixed point in the non-interacting limit requires the presence of a zero-energy Andreev bound state $f = (\gamma_1+i\gamma_2)/2$ at the junction as shown schematically in (b), with fine-tuning such that the wire couples \emph{only} to (say) $\gamma_1$.  (c) Flow diagram for the junction as a function of the interaction parameter $g$ for the Luttinger liquid.  For helical Luttinger liquids with $g < 2$, the perfect normal reflection fixed point is stable.  When $ g > 2$, however, this fixed point is unstable towards the Andreev fixed point.  Remarkably, here the Andreev bound state required to achieve perfect Andreev reflection will be generated dynamically, and with no fine-tuning required.}
\label{OrdinarySCjunction}
\end{figure}

We turn next to junctions formed by ordinary superconductors adjacent to helical Luttinger liquids, as sketched in Fig.\ \ref{OrdinarySCjunction}(a).  One may intuitively expect that here all traces of Majorana physics captured in the previous section are simply absent, since the superconductor no longer supports protected Majorana zero-modes.  While this is indeed true in the non-interacting limit, we will demonstrate that Majorana modes \emph{can} generically appear in the case of interacting helical Luttinger liquids.  To establish this counterintuitive result, we first note that in our analysis of generic properties of the S-matrix for free fermions in Sec.\ \ref{SmatrixProperties}, the topological nature of the superconductor we were considering there played an irrelevant role.  The same analysis [with the sole modification of dropping the $\phi_E \gamma_1$ term in Eq.\ (\ref{Gamma})] applied to present case leads to identical conclusions---in the $E \rightarrow 0$ limit the S-matrix must again be either purely diagonal or purely off-diagonal.  Thus even in the case of a junction formed with an ordinary superconductor, two physically allowed fixed points for free fermions remain, corresponding to perfect normal reflection and perfect Andreev reflection at the interface.  We will adopt a similar program to that followed in the previous section, beginning by understanding how to access these two fixed points for free fermions, and then addressing the stability of these fixed points when interactions are present using bosonization.

\subsection{Perfect normal and Andreev reflection fixed points for free fermions}
\label{3A}

Let us start by making the (physically reasonable) assumption that the ordinary superconductor is fully gapped throughout, so that the helical wire hosts the only low-energy degrees of freedom.  We again write the full Hamiltonian for the problem as $H = H_0 + \delta H$, where $H_0$ is the kinetic energy for the wire defined in Eq.\ (\ref{H0psi}) [or equivalently Eq.\ (\ref{H0})] and $\delta H$ contains the terms acting at the boundary.  Retaining only the leading potential and pairing terms at the junction, one has
\begin{equation}
  \delta H = \int _{-\infty}^\infty dx \left[2u\psi^\dagger \psi + \left(i \Delta\psi \partial_x \psi + h.c.\right)\right]\delta(x),
\end{equation}
which simply corresponds to Eq.\ (\ref{deltaH}) considered previously without the Majorana term.  The S-matrix for our Hamiltonian can therefore be read off from Eqs.\ (\ref{SmatrixAndreev}) by simply setting $\tilde t = 0$; this yields
\begin{eqnarray}
  S_{PP} &=& -\frac{(\tilde u+i)^2 + (\tilde E \tilde \Delta)^2}{1 +
     \tilde u^2 + (\tilde E \tilde \Delta)^2}
  \nonumber \\
   S_{PH} &=& \frac{2i \tilde E \tilde \Delta}{1 +
     \tilde u^2 + (\tilde E \tilde \Delta)^2},
   \label{SmatrixOrdinary}
 \end{eqnarray}
where again the tildes denote quantities normalized by $v_F$.

Notice that when $E \rightarrow 0$, $\Delta$ drops out and the S-matrix takes the form of Eq.\ (\ref{ReflectingBCs}) corresponding to perfect normal reflection.  No fine-tuning of parameters in $\delta H$ can alter this conclusion, nor can additional local perturbations involving only $\psi, \psi^\dagger$ (which necessarily carry additional derivatives than those already included above, and are thus unimportant at low energies).  Clearly then the perfect normal reflection fixed point is stable in the non-interacting limit.


How, then, can one access the physically allowed condition of perfect Andreev reflection?  For free fermions, accessing this fixed point requires the presence of an additional localized Andreev bound state in the superconductor, with corresponding operator $f$, to which the helical wire can couple; see Fig.\ \ref{OrdinarySCjunction}(b) for a schematic illustration.  (Our conclusions for the general properties of the S-matrix discussed in Sec.\ \ref{SmatrixProperties} again remain unaltered by the addition of this mode.)  To see how this transpires, let us write $f$ in terms of Majorana operators $\gamma_{1,2}$ via $f = (\gamma_1 + i \gamma_2)/2$ and consider the following interface Hamiltonian,
\begin{eqnarray}
  \delta H &=& \int_{-\infty}^\infty dx\left[\frac{t}{\sqrt{2}} \gamma_1(\psi^\dagger-\psi) + i \frac{t'}{\sqrt{2}}\gamma_2(\psi^\dagger + \psi)\right]\delta(x)
  \nonumber \\
  &+& i \frac{\delta}{2} \gamma_1\gamma_2
  \label{deltaH_with_gammas}
\end{eqnarray}
The $t, t'$ terms encode the most general bilinear couplings between $\psi,\psi^\dagger$ and $f,f^\dagger$ (after shifting these operators by overall phases to make the couplings real), while $\delta$ sets the energy for the Andreev bound state, which will generically be non-zero.  We have dropped the $u$ and $\Delta$ terms considered above for simplicity since these constitute qualitatively unimportant perturbations here.  With this form of $\delta H$ the S-matrix elements are now given by
\begin{eqnarray}
  S_{PP} &=& -\frac{(\tilde t \tilde t'-i \tilde \delta)^2 + \tilde E^2}{(\tilde t \tilde t')^2 + \tilde \delta^2 - i (\tilde t^2 + \tilde t'^2) \tilde E - \tilde E^2}
  \nonumber \\
  S_{PH} &=& \frac{-i (\tilde t^2-\tilde t'^2) \tilde E }{(\tilde t \tilde t')^2 + \tilde \delta^2 - i (\tilde t^2 + \tilde t'^2) \tilde E - \tilde E^2},
  \label{SmatrixWithAndreevBoundState}
\end{eqnarray}
with $\tilde t = t/v_F$, \emph{etc}.\ in our usual notation.
As before, one finds that in the limit $E\rightarrow 0$ the off-diagonal elements still generally vanish, reflecting stability of the perfect normal reflection fixed point even in the presence of an additional Andreev bound state at the junction.
One can, however, now fine-tune couplings in $\delta H$ to obtain the desired perfect Andreev reflection at zero energy by taking $\delta = 0$ and either $t = 0$ or $t' = 0$.  For concreteness let us choose $t' = \delta = 0$, upon which the S-matrix components become simply
\begin{eqnarray}
  S_{PP} &=& \frac{\tilde E}{i \tilde t^2  + \tilde E}
  \nonumber \\
  S_{PH} &=& \frac{i \tilde t^2 }{ i \tilde t^2 + \tilde E}.
\end{eqnarray}
Taking $E \rightarrow 0$ in this fine-tuned limit produces an S-matrix of the form in Eq.\ (\ref{AndreevBCs}), which indeed corresponds to perfect Andreev reflection.

More physically, in the non-interacting limit perfect Andreev reflection in the $E \rightarrow 0$ limit can emerge in this type of junction only when a pair of \emph{zero-energy} Majorana modes appears at the boundary, with the helical wire coupling to only \emph{one} of these.  Thus precisely as in the topological superconductor junction, the absorption of the Majorana mode $\gamma_1$ into the helical wire underlies the onset of perfect Andreev reflection.  One crucial difference of course is that this Majorana's partner, $\gamma_2$, is spatially separated in the topological case, which allows the Andreev fixed point to be stable for free fermions.  In the present context the coupling $t'$ to $\gamma_2$ in Eq.\ (\ref{deltaH_with_gammas}) is generically non-zero, and constitutes a relevant perturbation which drives the system back to the perfect normal reflection fixed point.

A second, and intimately related, difference relative to the topological case is that here $\gamma_1$ and $\gamma_2$ generally combine to form an Andreev bound state with finite energy $\delta$.  Interestingly, provided $t' = 0$ perfect Andreev reflection nevertheless survives when incident electrons in the helical wire are resonant with this bound state.  This can be understood from Eqs.\ (\ref{SmatrixWithAndreevBoundState}) by noting that when $t' = 0$ the S-matrix becomes purely off-diagonal at energy $E = \delta$.  Similar behavior has been captured previously by Law \emph{et al}.\cite{ZeroBiasAnomaly3} in the context of tunneling into discrete Majorana edge modes in a two-dimensional $p+ip$ superconductor.  However, we believe this behavior is special to the free Fermi case $g=1$ and does not survive in the interacting context.

Before turning to the interacting case it will prove beneficial to discuss the $t'$ and $\delta$ perturbations from a slightly different perspective.  At the perfect Andreev reflection fixed point, the coupling $t$---which must be treated non-perturbatively---strongly constrains the behavior of $\gamma_1$ at low energies.  More precisely, at energies $E \ll t$ the dynamics of $\gamma_1$ are `slaved' to those of $\psi$ and $\psi^\dagger$ such that the system avoids paying a large energy cost from the $t$ term in Eq.\ (\ref{deltaH_with_gammas}).  One can show this explicitly by diagonalizing the Hamiltonian $H = H_0 + \delta H$ with $t' = \delta = 0$,
\begin{equation}
  H = \int dx \left[-iv_F\psi^\dagger \partial_x \psi + \frac{t}{\sqrt{2}}\gamma_1(\psi^\dagger-\psi)\delta(x)\right],
  \label{H0_plus_t}
\end{equation}
for a finite-size wire and then expanding $\psi$ and $\gamma_1$ in terms of the resulting low-energy modes.  This calculation is sketched in Appendix \ref{MajoranaSolution}, and yields the familiar perfect Andreev reflection boundary condition, $\psi(x = 0^+) = \psi_R(x = 0) = \psi^\dagger(x = 0^-) = \psi_L^\dagger(x = 0)$, along with the following constraint on $\gamma_1$ at low energies:
\begin{equation}
  \gamma_1 = i \frac{v_F}{\sqrt{2}t}[\psi^\dagger(0^+)-\psi(0^+)-\psi^\dagger(0^-) + \psi(0^-)].
  \label{gamma_1_constraint}
\end{equation}

One can understand this result more intuitively in the following manner.  Let us decompose $\psi(x)$ in terms of two Majorana modes by writing
\begin{equation}
  \psi(x) = \frac{\gamma_r(x) + i \gamma_i(x)}{2};
\end{equation}
in this new basis the Hamiltonian reads
\begin{align}
    H &= -i\int dx\bigg{[}\frac{v_F}{4}(\gamma_r \partial_x\gamma_r + \gamma_i \partial_x \gamma_i)
  +\frac{t}{\sqrt{2}}\gamma_1\gamma_i\delta(x)\bigg{]}.
  \label{H_in_Majorana_basis}
\end{align}
While $\gamma_r(x)$ is unaffected by $\gamma_1$, the Majorana $\gamma_i(x)$ exhibits a sign change at $x = 0$ due to the coupling $t$.  Ultimately this sign change gives rise to the perfect Andreev reflection boundary conditions.  Since $\gamma_i(x)$ behaves singularly at the interface it is helpful to introduce a new Majorana operator $\tilde \gamma_i(x)$ which is well-behaved everywhere by defining
\begin{equation}
  \gamma_i(x) = {\rm sgn}(x) \tilde\gamma_i(x).
  \label{gamma_i_tilde}
\end{equation}
Differentiating the Euclidean action corresponding to Eq.\ (\ref{H_in_Majorana_basis}) with respect to $\gamma_i$ yields the following equation of motion:
\begin{equation}
  0 = (\partial_\tau-i v_F\partial_x)\gamma_i + i \sqrt{2}t\gamma_1 \delta(x).
\end{equation}
Rewriting this using Eq.\ (\ref{gamma_i_tilde}) yields a second term proportional to $\delta(x)$ due to the singular nature of the transformation at $x = 0$:
\begin{equation}
  0 = {\rm sgn}(x)(\partial_\tau \tilde \gamma_i - i v_F \partial_x \tilde \gamma_i)+i\sqrt{2}t\left(\gamma_1 - \frac{\sqrt{2}v_F}{t}\tilde\gamma_i\right)\delta(x).
\end{equation}
At energies $E \ll t$, the fields must conspire to eliminate the boundary terms above so that $t$ disappears entirely from the equation of motion.  This indeed occurs if we pin $\gamma_1 = \frac{\sqrt{2}v_F}{t}\tilde\gamma_i$, which agrees with Eq.\ (\ref{gamma_1_constraint}) derived by completely different means.

It follows from Eqs.\ (\ref{deltaH_with_gammas}) and (\ref{gamma_1_constraint}) that the two leading perturbations about the perfect Andreev reflection fixed point, given by turning on $t'$ and $\delta$, project to the same operator at low energies.

\subsection{Stability of the perfect normal and Andreev reflection fixed point with interactions}

Having understood how to access the physically allowed perfect normal and Andreev reflection fixed points for free fermions, we proceed now to assess the stability of each when interactions are present, beginning with the former.
Just as in Sec.\ \ref{NormalFixedPointStability}, bosonizing the Hamiltonian and imposing perfect normal reflection boundary conditions pins $\theta(x=0)$ to either 0 or $\pi$.  Equation (\ref{NormalFixedPointAction}), repeated here for clarity,
\begin{equation}
S_{\rm normal} = \frac{g}{2\pi} \int \frac{d\omega}{2\pi}|\omega| |\Phi|^2,
  \label{NormalFixedPointAction2}
\end{equation}
again describes the fixed point action for the fluctuating field $\Phi \equiv \phi(x = 0)$.  Due to the generic absence of Majorana modes at the junction, however, Eq.\ (\ref{deltaSt}) no long provides the leading perturbation away from this fixed point.  Instead, the most relevant perturbation which induces Andreev reflection at the junction corresponds to processes in which a Cooper pair hops from the helical Luttinger liquid into the ordinary superconductor:
\begin{eqnarray}
  \delta S_{\Delta} &=& \Delta \int d\tau [i\psi_R^\dagger(x = 0) \psi_L^\dagger(x = 0) + h.c.]
  \nonumber \\
  &\sim& 2\Delta \int d\tau \sin(2\Phi).
  \label{deltaSAndreev2}
\end{eqnarray}
Equation (\ref{deltaSAndreev2}) can be obtained via point-splitting in the unfolded chiral fermion theory: $\exp(2i\Phi)$ is the leading operator appearing in the operator product expansion of two $\psi^\dag$ fields.  Since $\sin(2\Phi)$ has scaling dimension $2/g$, the coupling $\Delta$ flows according to
\begin{equation}
  \frac{d\Delta}{d\ell} = (1-2 g^{-1})\Delta
\end{equation}
and is therefore irrelevant for $g < 2$.  Thus the perfect normal reflection fixed point is stable not only for free fermions, but also for helical Luttinger liquids with arbitrarily strong repulsive interactions, or attractive interactions below a critical strength.  When attractive interactions exceed this critical strength, resulting in $g > 2$, $\Delta$ then constitutes a relevant perturbation which drives the system away from this fixed point.  Physically, superconductivity is induced at the endpoint and spreads to the rest of the wire, resulting in a `topological' superconductor, but without exponentially localized end states.  Rather, because the superconductivity was seeded at only one endpoint, the Majorana modes both effectively live at that endpoint, with a splitting that turns out to be power law vanishing in the length of the wire.  We will now argue more precisely that the helical Luttinger liquid flows to perfect Andreev reflection, and explain the physical consequences.

Duality once again provides an effective tool for identifying the fate of the system when $g > 2$.  Following the steps outlined in Eqs.\ (\ref{Z}) through (\ref{Sdual}) to dualize $S_{\rm normal} + \delta S_{\Delta}$, one obtains the dual action
\begin{equation}
  S_{\rm dual}' = \int \frac{d\omega}{2\pi}\frac{|\omega|}{2\pi g}|\Theta|^2-v' \int d\tau \cos\Theta.
\end{equation}
Here $\Phi$ and $\Theta/\pi$ are once again conjugate variables, so that $\cos\Theta$ represents an instanton operator which tunnels between adjacent minima of the $\sin(2\Phi)$ potential in Eq.\ (\ref{deltaSAndreev2}).  Thus $v'$ is expected to be irrelevant when $\Delta$ is relevant (and vice versa), suggesting that $S_{\rm dual}'$ with $v' = 0$ describes the physically allowed perfect Andreev reflection fixed point that we identified in the free fermion case.  Furthermore, since $\cos\Theta$ represents the fermion parity in the helical Luttinger liquid, the dual action above also suggests the following: 1) at the fixed point with $v' = 0$ there is a degeneracy between states with even and odd fermion number, and 2) the leading perturbations away from this fixed point split this degeneracy.  We will now put these statements on firmer footing by explicitly constructing the fixed point action and leading perturbations beginning from the fermionic theory.

As in Sec.\ \ref{AndreevFixedPointStability}, the perfect Andreev reflection boundary condition pins $\phi(x = 0) = 0~{\rm or}~\pi$, and the fixed point action for $\Theta \equiv \theta(x = 0)$ derived by bosonizing the Hamiltonian and integrating out the fields away from $x = 0$ is given by
\begin{equation}
  S_{\rm Andreev} = \int \frac{d\omega}{2\pi}\frac{|\omega|}{2\pi g}|\Theta|^2,
  \label{AndreevFixedPointAction2}
\end{equation}
which indeed recovers $S_{\rm dual}'$ in the $v' = 0$ limit.  One might naively guess that the most relevant perturbation that induces normal reflection is the $u$ term in Eq.\ (\ref{deltaSnormal}), but this is incorrect.  To properly capture the physics it is crucial to recall how we accessed the Andreev fixed point in our scattering analysis for free fermions above.  Doing so required the presence of two zero-energy Majorana modes at the junction, $\gamma_1$ and $\gamma_2$ in Fig.\ \ref{OrdinarySCjunction}(b), along with fine-tuning such that the helical wire absorbed $\gamma_1$ (say) but decoupled completely from $\gamma_2$.  Without interactions perfect Andreev reflection at zero energy emerges \emph{only} in this fine-tuned limit.  One can then see that the leading perturbations correspond to the $\delta$ term in Eq.\ (\ref{deltaH_with_gammas}) which lifts the energy of $\gamma_{1,2}$ and the coupling between the Luttinger liquid and the Majorana mode $\gamma_2$,
\begin{eqnarray}
  \delta S_{\gamma_2} &=& i\frac{\delta}{2}\gamma_1\gamma_2
  \nonumber \\
  &+& \frac{t'}{\sqrt{2}}\int d\tau \gamma_2[\psi_R(x = 0)^\dagger + \psi_R(x = 0)].
\end{eqnarray}
Both perturbations were tacitly neglected by asserting that the system resided at the perfect Andreev reflection fixed point.  We have explicitly verified in Sec. \ref{3A} that both of these project to the same operator in the low energy limit, which corresponds to the one leading operator $\cos \Theta$ that appears upon performing the duality transformation.

Deducing the flow of $v'$ as discussed in Sec.\ \ref{NormalFixedPointStability}, one finds
\begin{equation}
  \frac{d v'}{d\ell} = [1-g/2]v'.
\end{equation}
For $g < 2$ this coupling is relevant and drives the system back to the perfect normal reflection fixed point, which is indeed stable in this regime.  More interestingly, for a helical Luttinger liquid with strong attractive interactions such that $g > 2$, coupling to this second Majorana mode represents an \emph{irrelevant} perturbation, implying stability of the perfect Andreev reflection fixed point.  In other words the fine tuning required to access perfect Andreev reflection for free fermions is no longer necessary with strong attractive interactions.  Figure \ref{OrdinarySCjunction}(c) summarizes the renormalization group flows found in this section for the ordinary superconductor/helical Luttinger liquid junction.  Similarly to the case of the helical/topological junction, when $g>2$ we expect one of the two dynamically generated Majorana modes to delocalize completely into the Luttinger liquid, and the other one to be power-law localized at the junction.

\section{Non-topological Superconductor---Spinful Luttinger liquid junctions}
\label{ntsll}

\begin{figure}
\includegraphics[width = 7cm]{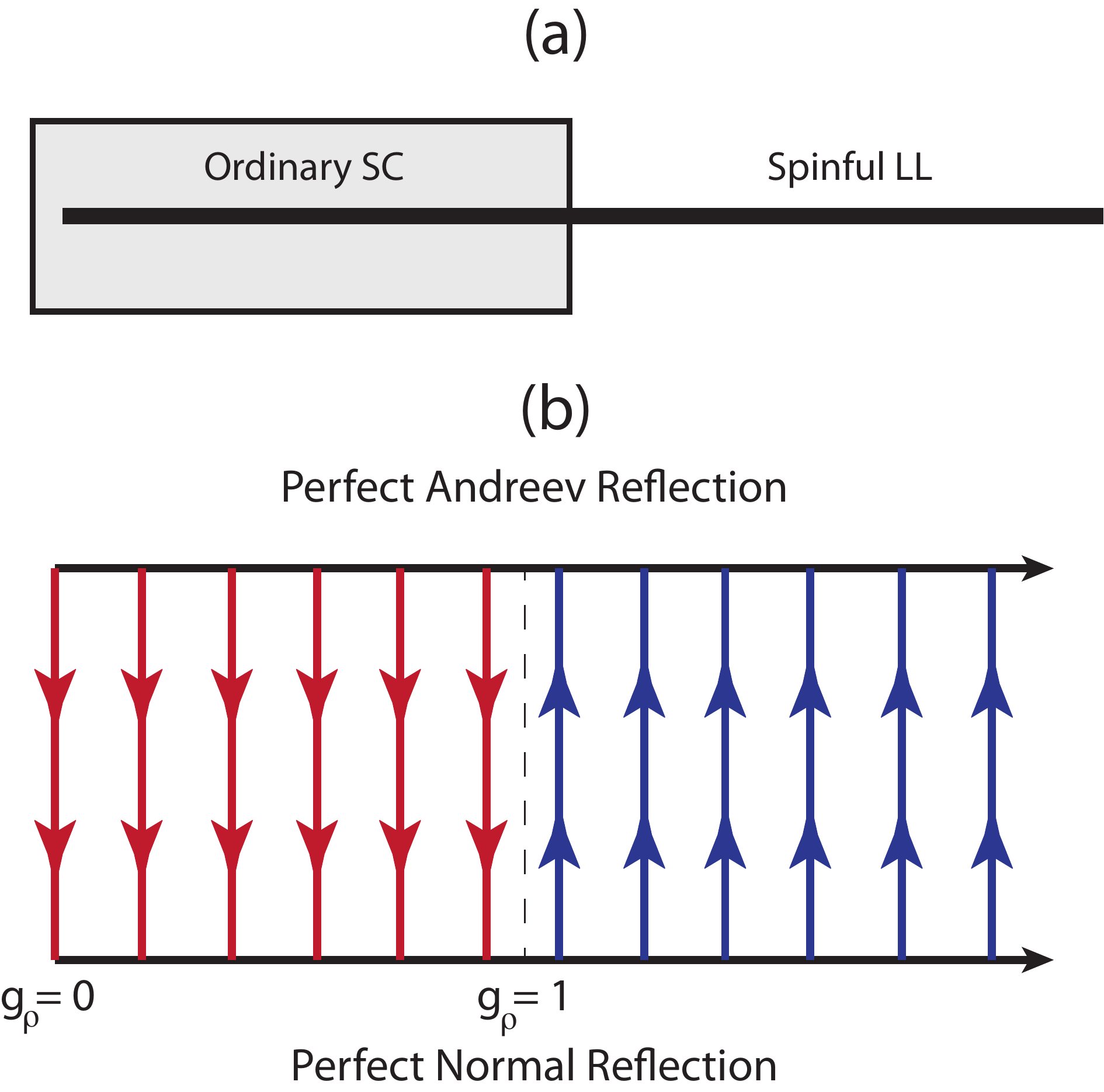}
\caption{(a) Trivial superconductor forming a junction with a spinful Luttinger liquid.  (b) Flow diagram for the junction as a function of the charge-sector interaction parameter $g_\rho$ for the Luttinger liquid, in the limit where the spin-sector interaction parameter is $g_\sigma = 1$.  Note that the free-fermion limit is very special---here there are marginal boundary couplings which lead to a non-universal zero-bias conductance ranging anywhere from $0$ to $4e^2/h$ depending on parameters.  For any repulsive interaction strength ($g_\rho < 1$), however, the junction flows to the perfect normal reflection fixed point where the zero-bias conductance vanishes.}
\label{OrdinarySCspinfulLL}
\end{figure}

We now consider a junction between a non-topological superconductor and a semi-infinite spinful Luttinger liquid [see Fig.\ \ref{OrdinarySCspinfulLL}(a)].  In previous work on this problem \cite{Maslov, LDOSdivergence}, conductance and local density of states were calculated at the Andreev fixed point of such a junction.  Here we emphasize that for weakly repulsive interactions such an Andreev fixed point is ultimately unstable, and the system generically crosses over to the normal reflecting fixed point.  This instability of the Andreev fixed point is in fact crucial for establishing sharp transport signatures of Majorana modes in the topological case considered in the next section.

We begin by bosonizing the problem.  The two spin channels $\alpha=\uparrow,\downarrow$ are bosonized according to
\be
\psi_{{R/L}\alpha} = e^{(i\phi_\alpha\pm i\theta_\alpha)}.
\ee
Defining the charge and spin fields
\begin{align}
\phi_{\rho/\sigma} &= \left(\phi_\uparrow \pm \phi_\downarrow\right) / \sqrt{2} \nonumber\\
\theta_{\rho/\sigma} &= \left(\theta_\uparrow \pm \theta_\downarrow\right) / \sqrt{2},
\end{align}
we then consider the general quadratic bulk action
\begin{align}
\label{bulk_action}
S_{\rm bulk} = \frac{v}{2\pi} \int_0^\infty dx \int d\tau & \bigg[ \frac{i}{\pi} \partial_x \theta_\rho \partial_\tau \theta_\rho + \frac{i}{\pi} \partial_x \phi_\sigma \partial_x \theta_\sigma \nonumber \\
&+ \frac{g_\rho}{2\pi}\left(\partial_x \phi_\rho\right)^2 + \frac{g_\sigma}{2\pi} \left(\partial_x \phi_\sigma\right)^2 \nonumber \\
&+ \frac{g_\rho^{-1}}{2\pi} \left(\partial_x \theta_\rho\right)^2 + \frac{g_\sigma^{-1}}{2\pi} \left(\partial_x \theta_\sigma \right)^2 \bigg].
\end{align}
Here $g_{\rho/\sigma}$ are the Luttinger parameters characterizing interactions in the charge/spin sectors, and $v$ the velocity.  To facilitate a Luttinger liquid analysis, we first (implicitly) assume a single velocity $v$ and also $g_\sigma=1$, which corresponds to a spin-$SU(2)$ invariant system.  We then discuss how the analysis is modified in the experimentally relevant situation where $SU(2)$ breaking terms are present.

There are two natural fixed points: normal reflecting, with $\theta_\rho, \theta_\sigma$ pinned at $x=0$, and Andreev reflecting, with $\phi_\rho$ and $\phi_\sigma$ pinned.  The boundary action for the normal fixed point is obtained by integrating out all fields except for $\phi_{\rho/\sigma}(x = 0)$, resulting in
\begin{align}
\label{spinful_norm}
S_{\rm normal} = \int \frac{d\omega}{2\pi} \frac{|\omega|}{2\pi} &\bigg[ \frac{(g_\rho+g_\sigma)}{2} \left(|\Phi_\uparrow|^2 + |\Phi_\downarrow|^2\right) \nonumber \\ &+ (g_\rho - g_\sigma) \Phi_\uparrow(\omega) \Phi_\downarrow(-\omega) \bigg],
\end{align}
where $\Phi_{\uparrow/\downarrow} \equiv \phi_{\uparrow/\downarrow}(x = 0)$.  The leading perturbation to this boundary action is the Cooper-pair tunneling term $\cos (\sqrt{2} \Phi_\rho)$, which has dimension $g_\rho^{-1}$ and is hence relevant for $g_\rho>1$.  In this range of $g_\rho$, the resulting Andreev fixed point to which the system then flows is simply the dual of Eq.\ (\ref{spinful_norm}):\footnote{We consider only bulk forward scattering interactions, which just renormalize the Luttinger liquid parameter.  Had we included $2k_F$ type interactions, other possibilities would have been allowed, such as an instability towards bulk s-wave superconductivity for $g_\rho>1.$}
\begin{align}
\label{spinful_Andreev}
S_{\rm Andreev} = \int \frac{d\omega}{2\pi} \frac{|\omega|}{2\pi} &\bigg[ \frac{(g_\rho^{-1}+g_\sigma^{-1})}{2} \left(|\Theta_\uparrow|^2 + |\Theta_\downarrow|^2\right) \nonumber \\ &+ (g_\rho^{-1} - g_\sigma^{-1}) \Theta_\uparrow(\omega) \Theta_\downarrow(-\omega) \bigg],
\end{align}
with $\Theta_{\uparrow/\downarrow} \equiv \theta_{\uparrow/\downarrow}(x = 0)$.  The leading perturbation is now normal backscattering, described by $\cos(\sqrt{2} \Theta_\rho)$, with scaling dimension $g_\rho$.

The key point is that in the physically relevant regime of weak repulsive interactions, $g_\rho < 1$, this normal backscattering term is relevant---\emph{i.e.}, the Andreev fixed point is \emph{unstable}.  Thus a junction between a spinful Luttinger liquid and an ordinary non-topological superconductor is described by a stable normal reflecting fixed point.  This stands in sharp contrast to the free fermion situation $g_\rho=1$, where these operators are exactly marginal and allow any value of the zero-bias conductance between $0$ and $4e^2/h$, as can be seen by explicitly solving the free fermion scattering problem.  (This is actually a special case of the free fermion solution for the topological/spinful junction presented in the next section; indeed, upon setting the coupling to the Majorana to zero one obtains the non-universal zero-bias conductance quoted here.)  Arbitrarily weak repulsive interactions drive a crossover to the normal reflecting fixed point, where the conductance goes to zero.  Figure \ref{OrdinarySCspinfulLL}(b) summarizes the renormalization group flows for this case.

Of course, in any proposed physical realization of a Majorana wire, spin-$SU(2)$ symmetry is broken by the Zeeman and spin orbit couplings (but in a simple Rashba model with density-density interactions and spin orbit coupling $g_\sigma=1$ because of a hidden $SU(2)$ symmetry 
\cite{Miles}).  At the level of free fermions, the only modifications these force on the low energy theory is differing Fermi momenta and velocities for the two species.  The former is simply a restriction on the types of operators that can appear as perturbations ({\it i.e.} some operators might not be allowed because of a $k_F$ mismatch), while the latter appears directly in the low energy action.  To include interactions we must bosonize, and it is easiest to do so in a basis which diagonalizes the velocity; that is, we separately bosonize the two spin modes with the differing velocities.  The kinetic term then becomes an arbitrary $2$ by $2$ symmetric matrix which generalizes the two Luttinger parameters $g_\rho$ and $g_\sigma$.  In principle we can integrate out the bulk to obtain a boundary theory and analyze the relevance of Cooper pair tunneling as above.  Although the resulting phase diagram depends in a complicated way on the three kinetic term coefficients and the two different velocities, it is still the case that the free fermion point is exactly marginal for any choice of velocities.  Hence there is a robust region of interaction parameter space near the free fermion fixed point - roughly speaking the set of repulsive interactions - where we are driven to the perfect normal reflecting fixed point.  Thus, even in the case of spin-$SU(2)$ symmetry breaking we generically expect no zero bias peak in the non-topological superconductor / spinful Luttinger liquid junction.

\section{Topological Superconductor---Spinful Luttinger liquid junctions}
\label{tsll}

\begin{figure}
\includegraphics[width = 7cm]{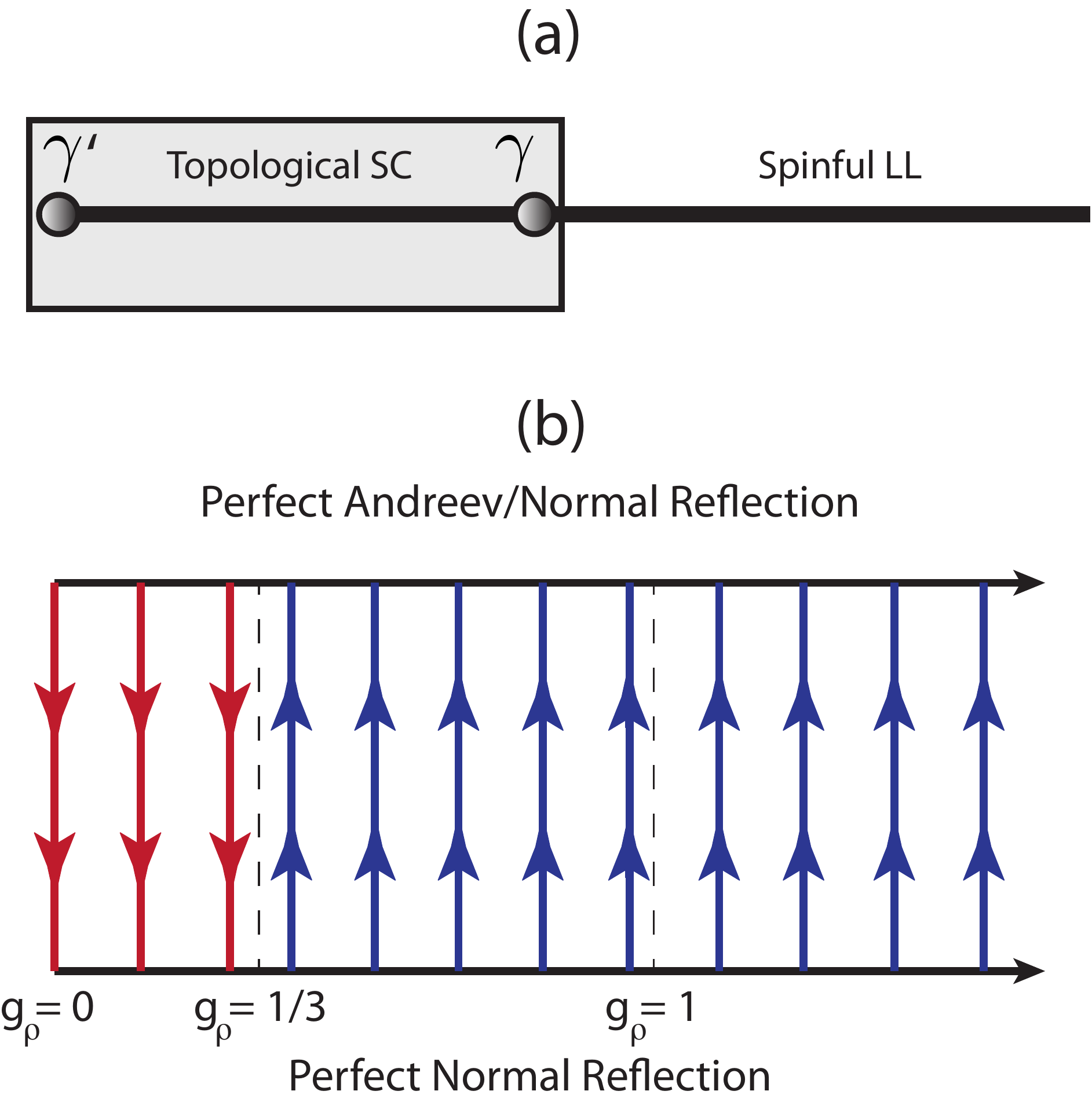}
\caption{(a) Topological superconductor forming a junction with a spinful Luttinger liquid.  (b) Flow diagram for the junction as a function of the charge-sector interaction parameter $g_\rho$ for the Luttinger liquid when the spin-sector interaction parameter is $g_\sigma = 1$.  For $g_\rho < 1/3$ the system flows to a perfect normal reflection fixed point characterized by a vanishing zero-bias conductance.  When $g_\rho > 1/3$, however, the junction flows to a novel fixed point corresponding to perfect Andreev reflection for one species and perfect normal reflection for the other, yielding a quantized $2e^2/h$ conductance. }
\label{TopoSCspinfulLL}
\end{figure}

Finally, we analyze the junction sketched in Fig.\ \ref{TopoSCspinfulLL}(a) between a topological superconductor and a semi-infinite spinful Luttinger liquid.  As in the helical case we will first attack the free-fermion limit and then treat the interacting case using bosonization.

\subsection{Scattering problem for free fermions}

We start with the free fermion scattering problem and compute the zero bias tunneling conductance when one allows for rather general boundary terms in the Hamiltonian.  Remarkably, we will show that this is robustly quantized at $2e^2/h$ independent of parameters, so long as coupling to the Majorana zero-mode at the junction remains finite.  Consider first the case when the spinful Luttinger liquid exhibits $SU(2)$ spin rotation symmetry in the bulk.  Coupling to the Majorana mode $\gamma$ at the junction can then always be gauge transformed to the form\cite{ZeroBiasAnomaly3}
\be
\delta H_\lambda = i \lambda \int dx \, \gamma \,(\psi_1 + \psi_1^\dag) \,\delta(x),
\ee
where $\psi_{1,2}$ can be obtained from $\psi_{\uparrow,\downarrow}$ via a rotation and a phase (we have replaced our previous label $\gamma_1$ with $\gamma$ for future notational clarity).  Neglecting terms involving derivatives, the most general quadratic form of the boundary Hamiltonian at $x=0$ reads
\begin{align} \label{generalL}
\delta H &= \int dx \,\delta(x) \, \big[ i\lambda\, \gamma \, (\psi_1+\psi_1^\dag) \nonumber \\
&+(\Delta_{12} \psi_1\psi_2 + v_{12}\psi_1^\dag \psi_2 + v_{11} \psi_1^\dag \psi_1 + v_{22} \psi_2^\dag \psi_2 + {\rm h.c.}) \big]
\end{align}
Because $\Delta_{12}$ and $v_{12}$ are complex while $v_{11}$ and $v_{22}$ are real, we have a total of $6$ real parameters.  The conductance follows upon computing the $S$-matrix as a function of these $6$ parameters and then summing up the probabilities of the Andreev processes.  

Let us now perform the calculation in detail.  We first decompose $\psi_{1,2}$ in terms of Majorana fermion field operators $\gamma_\mu$ as follows:
\begin{align}
\psi_1 &= \gamma_0 + i \gamma_2 \nonumber \\
\psi_2 &= \gamma_1 + i \gamma_3
\end{align}
and rewrite Eq.\ (\ref{generalL}) as
\be \label{Lreal}
\delta H = i \int dx \, \left[ \lambda \gamma \gamma_0(0) + \frac{1}{2} \sum_{\mu,\nu} g_{\mu\nu} \gamma_\mu(0) \gamma_\nu(0) \right].
\ee
Here $g_{\mu\nu}$ is a real anti-symmetric matrix, encoding the $6$ free parameters introducted earlier.  Let $S_0$ be the free fermion bulk action of the spinful wire and $\delta S$ be the boundary action corresponding to Eq.\ (\ref{Lreal}).  The equations of motion that follow from varying $S_0 + \delta S$ are:
\begin{align}
0&=\frac{\delta (S_0+\delta S)}{\delta\gamma}=i\lambda\gamma_0(0) - E \gamma \nonumber \\
0&=\frac{\delta (S_0+\delta S)}{\delta\gamma_\mu} = -i \partial_x \gamma_\mu + i\sum_\nu g_{\mu\nu} \gamma_\nu(0) \delta(x)
\nonumber \\
&- i\lambda\gamma\delta_{\mu 0}\delta(x) -E \gamma_\mu.
\end{align}
Here we have used time translation invariance to restrict to a solution with energy $E$.  To handle the derivatives and delta functions at $x=0$, we introduce the notation $\gamma_i(0^{\pm}) = \gamma_i(\pm \epsilon)$.  At $E=0$, the first equation of motion then gives $\gamma_0(0^+) = -\gamma_0(0^-)$, and thus the remaining ones become
\be
\gamma_i(0^+)=\gamma_i(0^-) + \sum_{j=1}^3 g_{ij} \gamma_j(0)
\ee
for $j=1,2,3$.  These can be summarized as
\be \label{invS}
\gamma_\mu(0^+) = \sum_\nu M_{\mu\nu} \gamma_\nu(0^-)
\ee
where $M_{00}=-1$, $M_{0i} = M_{i0}=0$, and $M_{ij} = A^{-1} B$, with $A_{ij} = \delta_{ij} - g_{ij}$, $B_{ij} = \delta_{ij} + g_{ij}$.  In fact, the $3\times3$ matrix $M_{ij}$ is just a generic element of $SO(3)$.  Equation (\ref{invS}) gives the scattering matrix in terms of Majorana fields.  Changing basis to complex Fermi fields and calculating the four Andreev transmission probabilities $\psi^\dag_{i} \rightarrow \psi_j$, we find that they generically add up to $1$; consequently the zero-bias conductance is indeed quantized at $2e^2/h$.

What about the physically relevant $SU(2)$ non-invariant case?  For gapless free fermions, the only way to break $SU(2)$ invariance is to introduce different velocities $v_{1,2}$ so that
\be
 H_0 = \int_0^\infty dx \sum_{\alpha = 1,2} \left[-iv_\alpha \left( \psi^\dag_{R\alpha} \partial_x \psi_{R\alpha} - i \psi^\dag_{L\alpha}\partial_x \psi_{L\alpha} \right)\right].
\ee
Now we cannot rotate the fields in such a way that only $\psi_1$ couples to the Majorana mode $\gamma$, and are forced to retain both $\lambda_0 \gamma \gamma_0(0)$ and $\lambda_1 \gamma \gamma_1(0)$.  Proceeding as in the $SU(2)$-invariant case we again derive Eq.\ (\ref{invS}), where now $M = {\tilde A}^{-1} {\tilde B}$.  The matrices on the right side are given by ${\tilde A}=V+G+L, {\tilde B} = V-G-L$, with $4\times4$ matrices $V, G, L$ defined as follows: $V = {\rm diag} (v_1, v_2, v_1, v_2)$, $L_{\mu\nu}=-\frac{i}{E} \lambda_\mu \lambda_\nu$ (where $\lambda_\mu=0$ for $\mu>1$), and $G_{\mu\nu}$ a general anti-symmetric $E$-independent matrix containing the quadratic boundary couplings.

In the case of differing velocities, the matrix $M$ defined above is no longer equivalent to the scattering matrix, and in particular is not unitary.  Indeed, the unitarity of the $S$-matrix follows from the conservation of probability current, which is proportional to $V$.  The correct $S$ matrix is then given by
\be
S = \sqrt{V} \, M \frac{1}{\sqrt{V}}
\ee
Rewriting this equation as
\be
S = \left(\frac{1}{\sqrt{V}} A \frac{1}{\sqrt{V}}\right)^{-1} \left(\frac{1}{\sqrt{V}} B \frac{1}{\sqrt{V}}\right)
\ee
and noting that
\be
\frac{1}{\sqrt{V}} A \frac{1}{\sqrt{V}} = \delta_{\mu\nu} + \frac{1}{\sqrt{V}}\left(G+L\right)\frac{1}{\sqrt{V}}
\ee
and similarly for $B$, we can now perform the rotation along $\mu,\nu=0,1$ to eliminate $\lambda_1$.  Taking the limit $E\rightarrow 0$ we recover the same form of the $S$-matrix as in the $SU(2)$ non-invariant case, and the same quantized conductance $G=2e^2/h$.

As remarked in the preceding section this result breaks down only when the coupling to the Majorana is fine-tuned to zero.  In this special limit one obtains a free fermion non-universal zero-bias conductance ranging from 0 to $4e^2/h$ [in contrast to the helical case, Pauli blocking is absent here so that the $\Delta_{12}$ term in Eq.\ (\ref{generalL}) can efficiently transmit Cooper pairs into the superconductor].  From this perspective it is somewhat curious that when the Majorana coupling is restored, $\Delta_{12}$ is unable to enhance the zero-bias conductance beyond $2e^2/h$.  Evidently one of the channels [$\psi_1$ in the $SU(2)$-invariant limit] hybridizes with the Majorana mode $\gamma$, leading to Andreev boundary conditions but also blocking transport of $\psi_1 \psi_2$ Cooper pairs.  We should emphasize that this result is specific to having only two conducting channels.

\subsection{Phase diagram with interactions}

We now analyze the interacting case.  We first bosonize as in the previous section, obtaining the bulk action in Eq.\ (\ref{bulk_action}).  Though we write expressions with general $g_\sigma$, in the stability analysis we assume $g_\sigma=1$, \emph{i.e.}, unbroken $SU(2)$ spin rotation symmetry in the bulk of the Luttinger liquid.  Then we proceed to discuss the physically relevant $SU(2)$ non-invariant case.  As for the helical/topological junction, we now need to include an additional effective spin-$1/2$ degree of freedom for the Majorana modes at the ends of the topological superconductor.  For simplicity we assume that the Majorana mode at the junction only couples to (say) the spin-up electron.  Tunneling onto the boundary Majorana mode bosonizes to
\be
\delta S_\lambda = \lambda \int d\tau \sigma^x \cos \Phi_\uparrow.
\ee
[As before we define $\Phi_{\uparrow/\downarrow} = \phi_{\uparrow/\downarrow}(x = 0)$ and $\Theta_{\uparrow/\downarrow} = \theta_{\uparrow/\downarrow}(x = 0)$].  Having already dealt with the subtleties of the spin-$1/2$ degree of freedom in Sec.\ \ref{TopoSC-HelicalLL}, we now arbitrarily fix the eigenvalue of $\sigma^x$ to $+1$ and drop $\sigma^x$ from the subsequent analysis.

We begin with the normal reflecting fixed point realized at $\lambda=0$, where $\Theta_{\alpha}=0$.  To determine its stability we have to compute the scaling dimension of $\cos \Phi_\uparrow$, which using Eq.\ (\ref{spinful_norm}) is $(g_\rho^{-1} + g_\sigma^{-1}) / 4$.  With $g_\sigma=1$, this term is relevant for $g_\rho> 1/3$.  In this case, the system flows to a novel Andreev/normal ($A\otimes N$) fixed point where $\Phi_\uparrow$ and $\Theta_\downarrow$ are simultaneously pinned.  Here the spin-up electrons exhibits perfect Andreev reflection while the spin-down electrons undergo perfect normal reflection.  Deriving the boundary field theory action for this new fixed point is somewhat subtle.  We again start with the bulk action (\ref{bulk_action}), but this time we integrate out $\phi_\uparrow$ and $\theta_\downarrow$, taking care to respect the boundary conditions $\phi_\uparrow(x=0) = \theta_\downarrow(x=0) = 0$ (handling these boundary conditions incorrectly results in a spurious total derivative term for the bulk that non-trivially modifies the boundary field theory, yielding the wrong scaling exponents).  This yields
\begin{align} \label{S_crossed_int}
S &= \nonumber \\
&\frac{1}{2\pi} \int dx d\tau \bigg[ \frac{2}{(g_\rho+g_\sigma)} (\partial_\mu \theta_\uparrow)^2 + \frac{2}{g_\rho^{-1}+g_\sigma^{-1}}(\partial_\mu \phi_\downarrow)^2 \bigg] \nonumber \\
&+\frac{i}{\pi} \int dx d\tau \left(\frac{g_\rho - g_\sigma}{g_\rho+g_\sigma}\right) \left(\partial_\tau \theta_\uparrow \partial_x \phi_\downarrow - \partial_x \theta_\uparrow \partial_\tau \phi_\downarrow \right).
\end{align}
Note that the second (Berry phase) term is a total derivative, and can be integrated to give
\be \label{S_Berry_term}
S_{\rm Berry} = -\frac{i}{\pi} \int d\tau \left(\frac{g_\rho-g_\sigma}{g_\rho+g_\sigma}\right) \Theta_\uparrow \partial_\tau \Phi_\downarrow
\ee
Integrating out $\phi_\downarrow(x)$ and $\theta_\uparrow(x)$ for $x>0$ from (\ref{S_crossed_int}) and combining the result with (\ref{S_Berry_term}) finally yields the following fixed-point action
\begin{align} \label{S_crossed}
&S_{A \otimes N}= \nonumber\\
&\,\,\,\,\,\, \int \frac{d\omega}{2\pi} \frac{|\omega|}{2\pi}\left[\frac{2}{(g_\rho^{-1} + g_\sigma^{-1})} |\Phi_\downarrow|^2 + \frac{2}{(g_\rho+g_\sigma)}|\Theta_\uparrow|^2 \right] \nonumber\\
&\,\,\,\,\,\,\,\,\,\,\,+\frac{i}{\pi} \left(\frac{g_\rho-g_\sigma}{g_\rho+g_\sigma}\right) \int d\tau\, \Phi_\downarrow \partial_\tau \Theta_\uparrow.
\end{align}
From Eq.\ (\ref{S_crossed}) we can determine all of the scaling dimensions.  In particular, $\psi_\uparrow \partial_x \psi_\uparrow \sim e^{2i\Theta_\uparrow}$ has scaling dimension $4/(g_\rho^{-1}+g_\sigma^{-1})$, and is thus relevant precisely for $g_\rho < 1/3$.  In that case the system flows back to the normal reflecting fixed point, whereas for $g_\rho>1/3$ it is the leading irrelevant operator around the new $A\otimes N$ fixed point.  These renormalization group flows are summarized in Fig.\ \ref{TopoSCspinfulLL}(b).

We now perform a more general analysis, involving all possible bulk perturbations - in particular, we include $SU(2)$ breaking terms.  Such bulk perturbations are important to analyze because they may qualitatively change the starting point for the boundary RG from which the $A \otimes N$ fixed point was derived.  At the free fermion fixed point, the leading physical perturbations are marginal: they involve various combinations of bilinears in $\partial \phi^{L/R}_\alpha$ and $e^{\pm i \phi^{L/R}_\alpha}$, where $\phi^{L/R}_\alpha$, $\alpha=1,2$ are the two left/right moving bosonic fields in an appropriately normalized basis.  Some of these operators correspond to changes in Luttinger parameters as well as perturbations away from the equal velocities case.  In the free fermion limit this does not open any bulk gaps, and we conjecture that this is the case even with interactions.  Hence they likely do not destabilize the boundary $A \otimes N$ fixed point.  There is also the operator $\cos (\sqrt{2} \theta_\sigma)$, which corresponds physically to an attractive $U$ Hubbard interaction inducing power-law superconducting long range order and the formation of a spin gap.  At this bulk RG fixed point electron tunneling onto the Majorana mode becomes highly irrelevant: the electrons are bound up in Cooper pairs.  Of course, the entire lead has then already become superconducting, so there is no meaningful way to discuss a zero bias anomaly due to Majorana modes in this case.  The remaining operators break the $SU(2)$ spin symmetry to a ${\mathbb Z}_2$ Ising symmetry which in turn is spontaneously broken, resulting in a ferromagnetic bulk.  This situation is dual to the one in which $\cos (\sqrt{2} \theta_\sigma)$ condenses, so the $A \otimes N$ fixed point does not survive here either.  However, for all physical $SU(2)$ breaking perturbations, such as spin orbit coupling and magnetic field, the $A \otimes N$ fixed point is stable.

\section{Discussion}
\label{sec:discussion}

In light of the intense current effort to realize Majorana fermions in condensed matter systems, it is important to understand the experimental signatures of these topologically protected zero modes.  In this work, we focused on tunnel junctions from normal Luttinger liquid leads to superconductors in order to probe the existence of Majorana zero modes in the latter.  Our approach was based on universality, and the low energy fixed points we found govern a wide variety of experimental setups---not just junctions in the limit of weak tunneling.  In particular, from a low energy universality point of view, it makes sense to consider four general junction archetypes, where in addition to the binary choice of topological or trivial superconductor one can also make the normal lead either helical (effectively spinless) or spinful.  We found that the transport signatures of such junctions are radically different and allow a robust distinction between topological and trivial superconductors.

Specifically, we found distinct low energy fixed points corresponding to perfect normal and perfect Andreev reflection, and determined their regimes of stability in the different junction archetypes.  Crucially, for the most physically relevant case of weak repulsive interactions we found that the stability of the normal reflecting fixed point is equivalent to the absence of a Majorana zero mode.  Conversely, the presence of a Majorana zero mode in this regime can be uniquely detected through a quantized conductance $G=2e^2/h$ characteristic of the Andreev fixed point.  We showed that such a quantized conductance is indeed a `smoking-gun' signature of topological superconductivity.  It should be again stressed that this distinction is not necessarily captured in the free fermion case---scattering theory in the non-interacting limit allows for similar conductances in a spinful/topological and spinful/non-topological junction.  Fortunately, repulsive interactions restore the sharp conductance dichotomy between these setups.

Let us summarize the physical consequences of the renormalization group flows for each of the junction archetypes.

\subsection{Helical wire---topological superconductor junction}

Let us first discuss the helical/topological junction, which in the regime $g>1/2$ should manifest a zero-bias anomaly associated with tunneling onto a Majorana fermion.  Typically captured in terms of free fermions \cite{ZeroBiasAnomaly0,ZeroBiasAnomaly1,ZeroBiasAnomaly2,ZeroBiasAnomaly3,ZeroBiasAnomaly4,ZeroBiasAnomaly5,ZeroBiasAnomaly6,ZeroBiasAnomaly7} (but see Refs.\ \onlinecite{MajoranaTeleportation,MajoranaCoulombBlockade}), we found the zero-bias anomaly to be robust to rather strong repulsive interactions and arbitrarily strong attractive interactions.  To calculate the value of the zero-bias conductance, we compute the imaginary time current-current correlation function
\begin{equation}
  \Pi(\tau) = \langle I(\tau)I(0)\rangle,
\end{equation}
which determines the conductance through the Kubo formula
\begin{equation}
  G = \frac{1}{\hbar}\frac{\Pi(i\omega \rightarrow \omega + i \delta)}{i\omega}.
\end{equation}
Since $\Theta/\pi$ gives the total charge on the Luttinger liquid, the current is $I = e \dot{\Theta}/\pi$.  Using the fixed-point boundary action of Eq.\ (\ref{Sdual}) (with $v=0$) it is straightforward to show that $\Pi(i\omega) = e^2 g|\omega|/\pi$, which apparently yields $G = g\left(2e^2/h\right)$.  However, this result fails to take into account the finite extent of the helical wire and the fact that it must ultimately be contacted by Fermi liquid leads [see Eq.\ (\ref{Sdual})].  This is a familiar problem in Luttinger liquid theory, and can be resolved in various ways by correctly modeling the wire together with the leads \cite{Maslov_Stone, Safi_Schulz}.  A particularly convenient scheme is to continuously interpolate the Luttinger parameter between its interacting value of $g$ and its free fermion value of $1$.  [One can implement this interpolation scheme with a spatially varying $g(x)$ that takes on a value of $g$ close to the junction and 1 far away.  In our boundary theory for the junction, this produces a frequency-dependent $g(\omega)$ that goes to one at zero frequency and $g$ at high frequency.]  This simple model already exhibits a crossover of the frequency dependent conductance $G(\omega)$ from $g\left(2e^2/h\right)$ for $\omega > v_F/L$ to $2e^2/h$ for $\omega < v_F/L$ (assuming this energy scale is smaller than the other relevant scales in the problem).  Hence, even in interacting junctions, the zero-bias conductance is expected to be quantized at $G= {2e^2}/{h}$.

\begin{figure}
\includegraphics[width = 7cm]{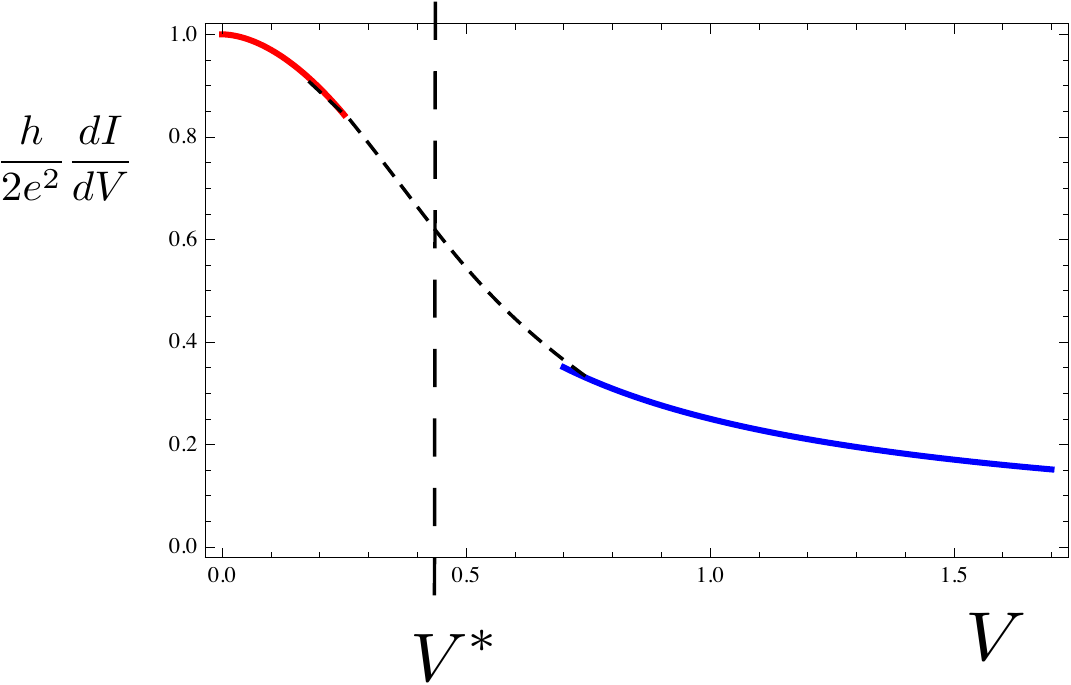}
\caption{Sample $dI/dV$ curve as a function of voltage $V$ (in arbitrary units) for $g=0.95$ in a helical nanowire/topological superconductor junction.  The blue and red portions are the perturbative results around the normal and Andreev fixed points respectively, calculated according to Eq.\ (\ref{scalingformG}).}
\label{scalingform}
\end{figure}

Universality also gives the line-shape of the finite bias conductance curve $dI/dV$ as a function of $V$.  The RG flow from the unstable normal reflecting fixed point to the stable Andreev one defines a crossover voltage $V^*$, which is roughly the width of the zero bias anomaly.  The entire $dI/dV$ curve can in principle be determined from universality.  Although the general calculation is involved, it is easy to calculate $dI/dV$ for $V \gg V^*$ and $V \ll V^*$ by performing perturbation theory around the normal and Andreev fixed points respectively.  We obtain
\begin{eqnarray}
  G \sim \left\{ \begin{array}{rl}
  (V/V^*)^{-2\left(1-1/(2g)\right)}, & V \gg V^* \\
 2e^2/h - (V/V^*)^{2(2g-1)}, & V \ll V^*,
       \end{array} \right.
   \label{scalingformG}
\end{eqnarray}
with the corresponding conductance line-shape illustrated in Fig.\ \ref{scalingform}.  Equation (\ref{scalingformG}) is valid down to voltage $\hbar v_F/(eL)$, which we assume to be much smaller than $V^*$.  Extracting the same value of $g$ from fitting experimental results to Eq.\ (\ref{scalingformG}) for both high and low $V$ would provide a non-trivial check of our results.

In fact, there is some subtlety involved in the derivation of (\ref{scalingformG}).  Indeed, for $V \gg V^*$, general scaling arguments show only that

\be G(V,t) = {\tilde G} \left(\frac{t/t_0}{(V/V_0)^{1-1/(2g)}}\right) \ee
where ${\tilde G}$ is a scaling function and $t_0$ is defined at the cutoff scale $V_0$.  ${\tilde G} (x)$ is an even function of $x$, so generically we expect ${\tilde G}(x) \sim x^2$.  However, for free fermions, (\ref{SmatrixAndreev}) and (\ref{difG}) show that ${\tilde G}(x) \sim x^4$, i.e. the coefficient of the quadratic term in ${\tilde G}$ vanishes.  We believe this to be highly non-generic, and expect that as soon as interactions are turned on, the coefficient becomes non-zero.  We can also consider finite temperature, in which case we have a two-parameter scaling function

\be
G(V,T,t) = {\doubletilde G} (  t/V^{1/2}, t/T^{1/2} )
\ee
Again, while in the free fermion case ${\tilde G}(x)$ has a non-generic vanishing of the quadratic term in $x$, the linear response conductance at high $T$ has a {\it non-vanishing} quadratic term.  That is, $\doubletilde G$ with the first argument set at infinity has non-zero second derivative with respect to the second term.  This qualitative distinction between zero temperature free Fermi and generic results again underscores the power of universality and the utility of our approach.

Apart from the conductance, the fixed points studied here can be further distinguished by the behavior of the local density of states at the junction.  Following Ref.\ \onlinecite{LDOSdivergence} the junction's local density states at frequency $\omega$ evaluated at each fixed point is given by
\begin{equation}
  \rho_{\rm LDOS}(\omega) \sim \left\{ \begin{array}{c}
  \omega^{1/g-1}~~~~\text{(perfect normal reflection)} \\
 \omega^{g-1} ~~~~  \text{(perfect Andreev reflection)}.
       \end{array} \right.
       \label{helicalLDOS}
\end{equation}
For the helical/topological junction, normal reflection is stable in the regime $g<1/2$, and the local density of states vanishes as a power law in $\omega$.  More interestingly, for $1/2 < g < 1$, which is likely the most physically accessible regime for solid state systems, the Andreev fixed point is stable and results in a \emph{divergent} density of states at low energies.  This remarkable feature has also been predicted for a spinful Luttinger liquid-ordinary superconductor junction with perfect Andreev reflection boundary conditions (but see the discussion below), and may be observable in tunneling measurements\cite{LDOSdivergence}.  Attractive interactions corresponding to $g > 1$ remove this divergence, despite the perfect Andreev reflection fixed point remaining stable there as well.

\subsection{Helical wire---non-topological superconductor junction}

\begin{figure}
\includegraphics[width = 7cm]{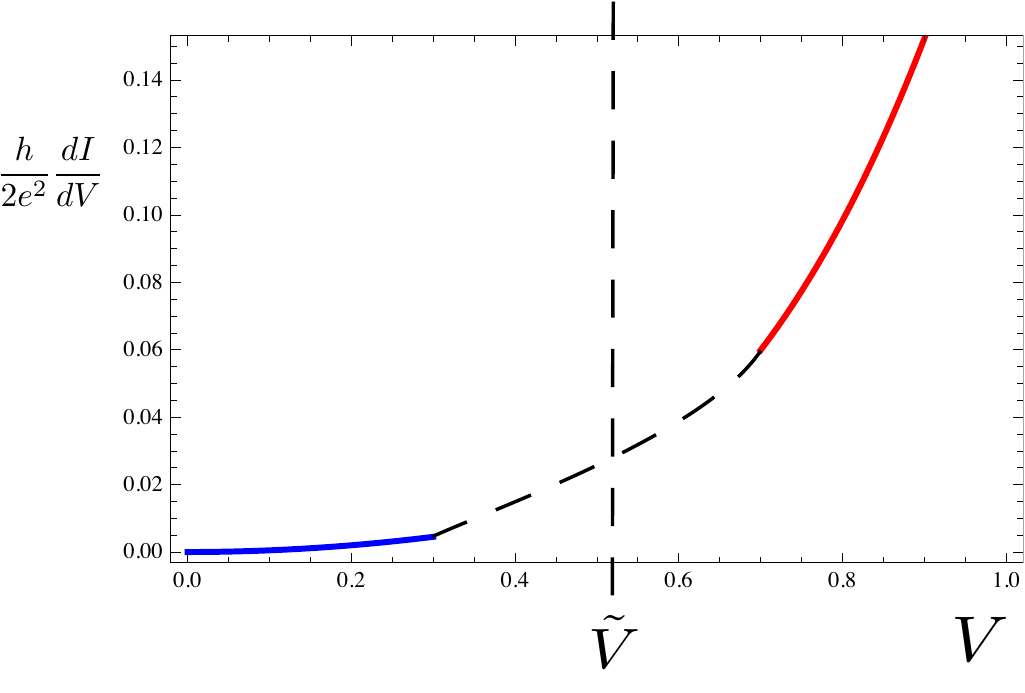}
\caption{Sample $dI/dV$ curve as a function of voltage $V$ (in arbitrary units) for $g=0.7$ in a helical nanowire/non-topological superconductor junction.  The red portion is calculated according to Eq.\ (\ref{scalingrep}), while at sufficiently low voltages (blue curve) the Fermi liquid leads dominate and we have a crossover to $g=1$ scaling.}
\label{scalingformrep}
\end{figure}

In the case of a helical/non-topological junction, the normal reflecting fixed point is stable for $g<2$.  Thus, in the physical regime of weak repulsive interactions, the zero-bias tunneling conductance is predicted to vanish.  We can again use perturbation theory to extract the finite bias conductance for small $V$.  We obtain the power law vanishing form
\begin{equation} \label{scalingrep}
G \sim (V/V^*)^{2(2/g-1)}.
\end{equation}
We expect this form to be valid down to voltages $e{\tilde V} \sim \hbar v_F/L$, below which point the finite length of the interacting Luttinger liquid becomes important.  Indeed, if we model the helical wire as a Luttinger liquid whose Luttinger parameter changes continuously from its interacting value of $g$ to the free Fermi value $1$ at $x\sim L$, then for $e{\tilde V} \sim \hbar v_F/L$ we are mostly sensitive to the long distance free Fermi part of the lead, and thus we expect the conductance to cross over to the behavior given by (\ref{scalingrep}) with $g=1$ (see Fig.\ \ref{scalingformrep}).

The stability of the normal reflecting fixed point in the physical regime $1/2<g<1$ also yields a different local density of states for the helical/non-topological junction.  Indeed, Eq.\ (\ref{helicalLDOS}) shows that
\begin{equation}
\rho_{\rm LDOS}(\omega) \sim \omega^{1/g-1}
\end{equation}
vanishes at zero energy.

\subsection{Spinful junctions}

The analysis of the spinful junctions proceeds analogously to that of the helical junctions.  In the case of a topological superconductor we once again have a zero-bias anomaly, and we can compute the shape of the conductance curve for high and low $V$ from the scaling dimensions of the leading perturbations to the normal and Andreev fixed points respectively.  Using the results of the previous section, and assuming $g_\sigma=1$ (\emph{i.e.}, spin $SU(2)$ invariance) for simplicity, we obtain
\begin{eqnarray}
  G \sim \left\{ \begin{array}{rl}
(V/V^*)^{-\frac{3-g_\rho^{-1}}{2}}, & V \gg V^* \\
 2e^2/h - (V/V^*)^{\frac{6g_\rho-2}{g_\rho+1}}, & V \ll V^*,
       \end{array} \right.
   \label{scalingformGspin}
\end{eqnarray}

The local density of states at the junction in the physically relevant regime of $g_\rho<1$ can be computed by evaluating the fermion two point function.  For the spinful/non-topological junction, we must evaluate the two point function at the normal reflecting fixed point, resulting in \be \rho_{\rm LDOS} (\omega) \sim \omega^{1/(2g_\rho)  - 1/2}, \ee
which goes to zero at low energy.  This is a generic feature of the $g_\rho<1$ regime of the spinful/non-topological junction, resulting from the instability of the perfect Andreev reflecting fixed point.  On the other hand, for the spinful/topological junction, the two point function must be evaluated at the Andreev fixed point, at least for $g_\rho>1/3$, resulting in \be \rho_{\rm LDOS} (\omega) \sim \omega^{g_\rho/2 - 1/2}. \ee  The local density of states thus diverges for weakly repulsive interactions in this case.

\subsection{Future directions}

Although we have shown that perfectly quantized conductance is a universal property of the low-energy/long-distance limit in topological junctions, a relevant issue for experiments is how closely one can approach this limit in practice.  In particular, in our analysis we have assumed a semi-infinite Luttinger liquid lead, and an arbitrarily long topological superconductor.  In a physical setup, neither assumption is valid.  In particular, a finite topological superconductor of length $L_{\rm SC}$ will have additional couplings $\delta$ and $t'$ (c.f. the discussion at the end of Sec. \ref{3A}) of magnitude $\sim\exp(-L_{\rm SC} / \xi)$, where $\xi$ is the induced coherence length.  These couplings are relevant (for $g<2$ in the helical case) and ultimately drive the system to the normal reflecting fixed point, with zero conductance.  However, if they are small to begin with, the crossover will occur at low energies and an intermediate regime with enhanced conductance is to be expected.

There exist many directions for further investigation.  One has to do with the nature of the crossover of the quantized conductance from $G=2e^2/h$ to $G = g(2e^2/h)$ as a function of the driving frequency $\omega$, which is expected to occur for $\omega \sim v_F/L_{\rm LL}$, where $L_{\rm LL}$ is the length of the Luttinger liquid lead.  Extracting the full dependence on both $\omega$ and the temperature $T$ is subtle and may require use of the Keldysh formalism.  Another is to study the fate of the Majorana zero mode for $g<1/2$ in the helical/topological junction.  Here we expect the Majorana mode to be hybridized with the degrees of freedom in the wire but nevertheless to remain power-law localized near the junction.  It would be interesting to study the form of this Majorana zero-mode using DMRG simulations.  Yet another direction would be to extend the present analysis to the case of the multi-channel case, which may be an experimentally relevant regime.  Another potentially experimentally relevant issue is the role of disorder in the Luttinger liquid lead, and its effect on the conductance.  Finally, it would be interesting to investigate multi-terminal junctions and possible universal signatures here.

\acknowledgments{It is a pleasure to thank Paul Fendley for stimulating discussions in early stages of this work.  We gratefully acknowledge support from the National Science Foundation through grants DMR-1055522 (J.A.), PHY-0803371 (N.H.L.), and DMR-1101912 (M.P.A.F.). M.P.A.F.\ also acknowledges funding provided by the Institute for Quantum Information and Matter at the California Institute of Technology, an NSF Physics Frontiers Center.  N.H.L.\ acknowledges support from the Gordon and Betty Moore Foundation through Caltech's Center for the Physics of Information.  We would like to thank the Aspen Center for Physics as well, where some of this work was initiated.
\appendix

\section{Derivation of the effective action for $\Phi$}

\label{PhiDeriv}

We start with the action
\be
S=\int d\tau \left( H(\phi,\theta) + \frac{i}{\pi} \int_0^\infty dx\,\partial_\tau \theta \partial_x \phi \right)
\ee
with
\be
H=\int_0^\infty dx \frac{v}{2\pi}\left[g(\partial_x\phi)^2+g^{-1}(\partial_x\theta)^2\right]
\ee
The normal reflection fixed point pins $\theta(x=0)$. Consider the right and left moving fields:
\be
\phi_R=\phi+\theta/g, \qquad \phi_L=\phi-\theta/g
\ee
The boundary condition $\theta(x=0)=0$ gives
\be
\phi_L(x=0)=\phi_R(x=0).
\ee
We define a new field $\tp(x)$ as
\bea
\tp(x)&=&\phi_R(x), \qquad x>0 \\
\tp(-x)&=&\phi_L(-x), \qquad x<0 \\
\eea
Note that $\tp(0)=\phi_R(0)=\phi_L(0)=\phi(0)$.  We have
\be
S_0= g\int dx d\tau \left[\frac{ v}{4\pi}(\partial_x
\tp)^2+\frac{i}{4\pi}\partial_\tau \tp \partial_x \tp \right]
\ee

We introduce a variable $\Phi(\tau)$ and a Lagrange multiplier $\Lambda$ as
\be
S_\Lambda=i\int d\tau \Lambda(\tau)(\Phi(\tau)-\tp(x=0,\tau))
\ee
We now integrate over the modes $\tp_{k,\omega}$. Note that
\be
S_0= \int \frac{dk \,d\omega}{(2\pi)^2} G^{-1}(k,\omega)
\tp_{k,\omega}  \tp_{-k,-\omega}
\ee
with
\be
 G^{-1}(k,\omega)=\frac{gv}{4\pi } \left[(k +i \frac{
\omega}{2v} )^2 + \frac{\omega^2}{4v^2}\right]
\ee
Performing the gaussian integral over $\tp_{k,\omega}$ yields
\be
S=i \int \frac{d\omega}{2\pi} \Lambda_\omega\Phi_{-\omega}+ \int
\frac{dk \,d\omega}{(2\pi)^2}\,
\frac{1}{4}\, G(k,\omega)
\Lambda_\omega \Lambda_{-\omega}
\ee
Performing the integral over $k$ gives
\be
S=\int \frac{d\omega}{2\pi} \left[
i\Lambda_\omega\Phi_{-\omega}+
\frac{ \pi }{ 2 g|\omega|} \Lambda_\omega \Lambda_{-\omega}\right]
\ee
Finally, performing the functional integral over $\Lambda_\omega$ yields
\be
S_\lambda=\int \frac{d\omega}{2\pi}  \frac{g|\omega|}{2\pi}
\Phi_\omega
\Phi_{-\omega}
\ee
as desired.

\section{Derivation of the path integral at the perfect normal reflection fixed point}
\label{PathIntregralDerivation}

Here we will derive the path integral representation for the partition function in Eq.\ (\ref{Z}), beginning from the full action that includes bulk degrees of freedom in the helical Luttinger liquid.  This is done by inserting resolutions of the identity between intermediate imaginary time steps, using the dual orthonormal bases
\begin{equation}
{\bf 1} = \sum_{s^x, \phi} |s^x\,\phi\rangle\langle s^x\,\phi| = \sum_{s^z,\theta} |s^z\,\theta\rangle\langle s^z\,\theta|.
\end{equation}
Here $s^{x,z}$ denote the eigenvalues $(=\pm1)$ of the corresponding Pauli operators $\sigma^{x,z}$, and $\phi, \theta$ are shorthand for eigenvalues of $\phi(x), \theta(x), 0 \leq x \leq L$.  To obtain the partition function we need the overlaps between these states.  Using
\begin{align}
|s^x=\pm 1\rangle &= \frac{1}{\sqrt{2}}\left(|s^z=1\rangle \pm |s^z=-1\rangle\right)
\end{align}
we obtain
\begin{equation}
\langle s^x|s^z\rangle = \frac{1}{\sqrt{2}} \exp\left[ i\pi \frac{(1-s^x)}{2} \frac{(1-s^z)}{2} \right]
\end{equation}
which together with the fact that $\phi(x)$ and $\partial_x \theta(x) / \pi$ are dual variables yields (up to unimportant constants)
\begin{align}
\langle & \phi\,s^x|\theta\,s^z\rangle \nonumber \\ &= \exp\left[ i \pi \frac{(1-s^x)}{2} \frac{(1-s^z)}{2} + \frac{i}{\pi} \int_0^L dx\,\phi \,\partial_x \theta \right].
\end{align}
It follows that the partition function can be expressed as
\begin{align}
Z = \int \mathcal{D}\phi\mathcal{D}\theta\sum_{\{s^x(\tau) \in \pm 1\} }\sum_{\{s^z(\tau) \in \pm 1\} } e^{-S_E+iS_B}
\label{Zeq}
\end{align}
with $S_E$ the usual Euclidean action
\begin{align}
S_E &= \frac{v}{2\pi} \int dx \,d\tau \left[\frac{1}{g}(\partial_x \theta)^2+ g(\partial_x \phi)^2 \right] \nonumber \\ &+ 2t \int d\tau \cos \left[\phi(x=0) + \pi\frac{1-s^x}{2} \right]
\end{align}
and $S_B$ the Berry phase contribution
\begin{equation}
S_B = \frac{\pi}{4} \int d\tau (1-s^z) \partial_\tau s^x + \frac{1}{\pi} \int dx \, d\tau\, \partial_\tau \phi \,\partial_x \theta. \label{S_Berry}
\end{equation}

Suppose that we sum over $s^x$ in Eq.\ (\ref{Zeq}).  To do so it is convenient to first rewrite the path integral exchanging $\phi(x,\tau)$ for the new variable
\begin{equation}
{\tilde\phi}(x,\tau) = \phi(x,\tau) + \frac{\pi}{2} \left(1-s^x(\tau)\right).
\end{equation}
Integration by parts shows that $s^x$ then appears only in the following contribution to the action:
\begin{equation}
\int d\tau \frac{(1-s^x)}{2} \partial_\tau\left[-\theta(x=0) + \frac{\pi}{2} \left(1-s^z\right) \right].
\end{equation}
Summing over $s^x$, at each discrete time step, gives a vanishing contribution to the partition function unless $\partial_\tau\left[-\theta(x=0) + \pi/2 \left(1-s^z\right) \right] = 0$ modulo $2\pi$ at each $\tau$.  Hence the exponential of $2\pi i$ times this quantity is conserved.  Converting to operator notation, this is $\sigma^z(\tau) \exp\left[i \theta(0,\tau)\right]$---which is just the fermionic parity of the superconductor plus Luttinger liquid system.

Let us now instead sum over $s^z$ in Eq.\ (\ref{Zeq}).  Using Eq.\ (\ref{S_Berry}), it follows that $s^x$ is a conserved quantity, so the path integral sums over only two imaginary time configurations of $s^x$: $s^x=+1$ for all $\tau$ and $s^x=-1$ for all $\tau$.  Integrating out everything but $\Phi = \phi(x=0)$ yields the action $S_{\rm normal} + \delta S_{t}$ derived above [see Eqs.\ (\ref{NormalFixedPointAction}, \ref{dstfinal})].  The partition function then becomes
\begin{eqnarray}
  Z &=& \int \mathcal{D} \Phi \sum_{s^x = \pm 1}e^{-S_{\rm normal}}e^{-2t \int d\tau s^x\cos\Phi},
  \label{Zrepeated}
\end{eqnarray}
which indeed recovers Eq.\ (\ref{Z}).

\section{Solution of a non-interacting helical wire coupled to a single Majorana mode}
\label{MajoranaSolution}

In this Appendix we will sketch the solution of the following Hamiltonian,
\begin{equation}
  H = \int_{-L}^L dx \left[-iv_F\psi^\dagger \partial_x \psi + \frac{t}{\sqrt{2}}\gamma_1(\psi^\dagger-\psi)\delta(x)\right],
  \label{H0_plus_t_repeated}
\end{equation}
which describes a non-interacting helical wire of length $L$ coupled to a single Majorana mode $\gamma_1$ at one end.  This Majorana's `partner' $\gamma_2$ is assumed to decouple entirely from both the wire and $\gamma_1$.  At the other end of the wire we will impose perfect normal reflecting boundary conditions, requiring
\begin{equation}
  \psi(x = L) = \psi(x = -L).
  \label{bc_at_L}
\end{equation}
Our goal will be to find the low-energy wavefunctions of $H$ in the limit $E \ll t$ and expand $\psi$ and $\gamma_1$ in terms of the corresponding modes.

As a first step we write $H$ in terms of $\Psi^\dagger(x) = [\psi^\dagger(x) \psi(x) f^\dagger f]$, where $f = (\gamma_1 + i \gamma_2)/2$:
\begin{eqnarray}
  H &=& \frac{1}{2}\int_{-L}^L dx \Psi^\dagger \mathcal{H}\Psi.
  \\
  \mathcal{H} &= &\left[ \begin{array}{cccc}
-i v_F \partial_x & 0 & -\frac{t}{\sqrt{2}}\delta(x) & -\frac{t}{\sqrt{2}}\delta(x) \\
0 & -i v_F \partial_x & \frac{t}{\sqrt{2}}\delta(x) & \frac{t}{\sqrt{2}}\delta(x) \\
-\frac{t}{\sqrt{2}}\delta(x) & \frac{t}{\sqrt{2}}\delta(x) & 0 & 0 \\
-\frac{t}{\sqrt{2}}\delta(x) & \frac{t}{\sqrt{2}}\delta(x) & 0 & 0 \end{array} \right].
\end{eqnarray}
The wavefunctions $\Phi_E(x)$ with energy $E$ can be immediately written for $x \neq 0$ as
\begin{eqnarray}
  \Phi_E(x>0) = \left[ \begin{array}{c}
e^{i \frac{E x}{v_F}}a_E^> \\ e^{i \frac{E x}{v_F}}b_E^> \\ c_E \\ d_E \end{array} \right], \Phi_E(x<0) = \left[ \begin{array}{c}
e^{i \frac{E x}{v_F}}a_E^< \\ e^{i \frac{E x}{v_F}}b_E^< \\ c_E \\ d_E \end{array} \right].
\end{eqnarray}
The elements above are constrained by normalization, the boundary condition of Eq.\ (\ref{bc_at_L}) which requires
\begin{eqnarray}
  e^{i \frac{EL}{v_F}}a_E^> &=& e^{-i \frac{EL}{v_F}}a_E^<
  \nonumber \\
  e^{i \frac{EL}{v_F}}b_E^> &=& e^{-i \frac{EL}{v_F}}b_E^<,
  \label{bc_constraints}
\end{eqnarray}
and the following relations needed to satisfy the Hamiltonian at $x = 0$,
\begin{eqnarray}
  0 &=& (a_E^>-a_E^<)+(b_E^>-b_E^<)
  \nonumber \\
  0 &=& i v_F(a_E^>-a_E^<) + \frac{t}{\sqrt{2}}(c_E+d_E)
  \label{constraints} \\
  E c_E &=& E d_E = \frac{t}{\sqrt{2}}( b_E^>- a_E^<).
  \nonumber
\end{eqnarray}
Equations (\ref{bc_constraints}) and (\ref{constraints}) admit a non-trivial solution provided the energies satisfy
\begin{equation}
  0 = \frac{E v_F}{t^2}\left[\cos\left(\frac{2 E L}{v_F}\right)-1\right] + \sin\left(\frac{2 E L}{v_F}\right).
\end{equation}
For $E \ll t$ the energies are well-approximated by
\begin{equation}
  E_n = \frac{n\pi v_F}{2L}
\end{equation} for integer $n$; the associated wavefunction components can be found to leading order in $E_n/t$ from Eqs.\ (\ref{bc_constraints}) and (\ref{constraints}).

One can deduce by inspection that one of the zero-energy wavefunctions supported by $H$ is
\begin{equation}
  \Phi_{\gamma_2} = \frac{1}{\sqrt{2}}\left[ \begin{array}{c}
0 \\ 0 \\ i \\ -i \end{array} \right],
\end{equation}
which simply corresponds to the decoupled Majorana mode $\gamma_2$.  The wavefunctions carrying energy $E_n$ with $n$ even are given by
\begin{equation} \label{b11}
  \Phi_{E_n} = \frac{1}{\sqrt{N_n}}\left[ \begin{array}{c}
e^{i \frac{E_n x}{v_F}} \\ e^{i \frac{E_n x}{v_F}} \\ 0 \\ 0 \end{array} \right]~~~(n ~{\rm even}),
\end{equation}
where $N_n$ is the normalization.  Note that $\Phi_{E_0}$ corresponds to the zero-energy Majorana mode which is absorbed into the helical wire due to the tunneling $t$.  Finally, the $n$ odd wavefunctions read
\begin{equation}
  \Phi_{E_n} = \frac{1}{\sqrt{N_n}}\left[ \begin{array}{c}
i{\rm sgn}(x)e^{i \frac{E_n x}{v_F}} \\ -i{\rm sgn}(x)e^{i \frac{E_n x}{v_F}} \\ \frac{\sqrt{2} v_F}{t} \\ \frac{\sqrt{2} v_F}{t} \end{array} \right]~~~(n ~{\rm odd}).
\end{equation}

With these wavefunctions in hand one can now expand $\psi(x)$ and $\gamma_1$ in terms of low-energy modes for the system:
\begin{eqnarray}
  \psi(x) &\sim & \sum_{n~{\rm even}} \frac{e^{i \frac{E_n x}{v_F}}}{\sqrt{N_n}} \Gamma_n + i {\rm sgn}(x)\sum_{n~ {\rm odd}}\frac{e^{i \frac{E_n x}{v_F}}}{\sqrt{N_n}} \Gamma_n'
  \nonumber \\
  \gamma_1 &=& f^\dagger + f \sim \frac{2\sqrt{2}v_F}{t}\sum_{n~{\rm odd}} \frac{1}{\sqrt{N_n}}\Gamma_n'.
  \label{expansion}
\end{eqnarray}
Here $\Gamma_n^\dagger = \Gamma_{-n}$ and $\Gamma_n'^\dagger = \Gamma_{-n}'$ respectively create energy $E_n$ excitations with $n$ even and odd.  Equations (\ref{expansion}) encode two important relations.  First, it follows that at low energies
\begin{equation}
  \psi(x = 0^+) = \psi_R(x = 0) = \psi^\dagger(x = 0^-) = \psi_L^\dagger(x = 0),
  \label{derived_Andreev_bc}
\end{equation}
which is the familiar perfect Andreev reflection boundary condition induced by the coupling to $\gamma_1$.  Second, $\gamma_1$ and $\psi, \psi^\dagger$ are not independent at low energies; using Eq.\ (\ref{derived_Andreev_bc}) their relation can be expressed in the following symmetric form
\begin{eqnarray}
  \gamma_1 &\sim& i \frac{v_F}{\sqrt{2}t}[\psi^\dagger(0^+)-\psi(0^+)-\psi^\dagger(0^-)+\psi(0^-)]
  \nonumber \\
  &=& i \frac{v_F}{\sqrt{2}t}[\psi_R^\dagger(0)-\psi_R(0)-\psi_L^\dagger(0)+\psi_L(0)].
\end{eqnarray}

\section{Solution of an interacting helical wire coupled to a single Majorana mode}
\label{interactingMS}

We now treat the interacting helical wire.  We first redo the calculation for the non-interacting case directly in terms of the bosonic modes, and then generalize to $g \neq 1$.  It is useful to bosonize both the Luttinger liquid and the two-level system formed by $\gamma_1,\gamma_2$, as done in Section \ref{NormalFixedPointStability}:
\begin{eqnarray} \gamma_1 &=& \sigma^y \\ \gamma_2 &=& \sigma^x. \end{eqnarray}
Also, because the Jordan-Wigner string goes to the left in ({\ref{JordanWigner}), the bosonized form of the continuum Fermi fields includes an extra factor of $\sigma^z$, which is simply the Fermionic parity of the $\gamma_{1}, \gamma_{2}$ system.  The tunneling term is $H_t = 2t \sigma^x \cos(\phi(x=0))$, but for convenience in this Appendix we shift the phase of $\phi(x)$ by $\pi/2$, so that the tunneling term becomes \begin{equation} H_t = 2t \sigma^x \sin(\phi(x=0)). \end{equation}  From now on we work in the low energy Hilbert space $\HHlow \subset \HH$ where $\phi(x=0)$ is pinned at $\pm \pi/2$, i.e. we assume we are exactly at the Andreev fixed point.  The state of the spin-$1/2$ representing the topological superconductor is then completely determined by $\phi(x=0)$, so that we need only retain the Luttinger liquid degrees of freedom to describe all the states and operators in $\HHlow$.

In particular, the operator $\gamma_2=\sigma^x$, representing the decoupled Majorana mode, is given by $\sin(\phi(x=0))$.  Now, if the fermionic Hamiltonian were quadratic, we would necessarily have a partner Majorana mode for $\gamma_2$, i.e. another fermionic operator which commuted with the Hamiltonian and squared to $1$.  Indeed, this operator is a zero momentum ``plane-wave" solution of the Bogoliubov-de Gennes equation (see Eq. \ref{b11}).  What about the interacting case?  In this section we derive a bosonized expression for the partner Majorana mode in such an interacting helical wire, for all $g>1/2$.  Furthermore, we check that this expression is correct perturbatively to leading order in $g-1$.

We begin by redoing the calculation for $g=1$ in the bosonic framework, and then generalize to $g \neq 1$.  The normal reflecting boundary condition at $\theta(L)=0$ is equivalent to $\frac{\partial \phi(L)}{\partial x}=0$.  A general $\phi$ field configuration can thus be expanded as \begin{equation} \label{pe} \phi(x) = \pm \pi/2+ \sum_{n=0}^{\infty} \phi_n \sin \frac{(2n+1)\pi x}{2L}. \end{equation}  The Hamiltonian for $g=1$ reads \begin{equation} \label{freeH} H = \frac{v_F}{2 \pi} \int_0^L dx \left[ \left(\partial_x \phi\right)^2 - \left(\pi \frac{\partial}{\partial \phi(x)}\right)^2 \right]. \end{equation}  Expanding \begin{equation} \frac{\partial}{\partial \phi_n} = \int_0^L dx \sin \left(\frac{(2n+1)\pi x}{2L} \right) \frac{\partial}{\partial \phi(x)} \end{equation} and inverting, we obtain \begin{equation} \label{bosonic_H} H = \frac{\pi v_F}{L} \sum_{n\geq 0} \left[ -\frac{1}{2} \left(\frac{\partial}{\partial \phi_n}\right)^2 + \frac{\omega_n^2}{2} \phi_n^2 \right] \end{equation} with $\omega_n = (2n+1)/4$.  We define creation and annihilation operators \begin{eqnarray} \phi_n &=& \frac{1}{\sqrt{2\omega_n}} \left( a_n + a_n^\dag \right) \\ \frac{\partial}{\partial \phi_n} &=& \sqrt{\frac{\omega_n}{2}} \left(a_n - a_n^\dag\right), \end{eqnarray} and expand out the field operators $\phi(x)$ and \begin{equation} \theta'(x) = -\pi i \int_x^L dy \frac{\partial}{\partial \phi(y)}. \end{equation}  Here $\theta'(x) = \theta(x) - \theta(L)$ sends the Jordan-Wigner string to the right, and thus preserves the boundary condition $\phi(0) = \pm \pi/2$.  We obtain: \begin{align} \label{phi}\phi(x) &= \pm \frac{\pi}{2} + \frac{i}{2} \sum_n \frac{e^{-i k_n x} - e^{i k_n x}}{\sqrt{2\omega_n}} \left(a_n^\dag + a_n\right) \\ \label{theta} \theta'(x) &= \frac{i}{2} \sum_n \frac{e^{-i k_n x} + e^{i k_n x}}{\sqrt{2\omega_n}} \left(a_n^\dag - a_n\right), \end{align} where $k_n = 2\pi\omega_n/L$. We can exchange $\theta$ for $\theta'$ at the expense of introducing an extra factor of $P_{LL} = \exp(\theta(0)-\theta(L))$, so that the bosonization now becomes \begin{eqnarray} \psi_{R/L} &\sim& \left(1-2b_{-1}^\dag b_{-1}\right) e^{i(\phi \pm \theta)} \nonumber \\ &=& P_{LL} e^{i(\phi \pm \theta')}. \end{eqnarray}
For the sake of efficiency, we now change notation $(\psi_R(x), \psi_L(x)) \rightarrow (r(x),\ell(x))$.  For a quadratic Fermionic Hamiltonian (g=1), one can easily verify that \be \label{quad_delta_1} \delta_1 = -\frac{i}{4L} \int_0^L dx \left(r-r^\dag+\ell-\ell^\dag \right) \ee
is the partner ``plane-wave" Majorana mode for $\gamma_2$.  $\delta_1$ is the result of $\gamma_1$ leaking into the Luttinger liquid via the coupling $t$, with (\ref{quad_delta_1}) valid in the low energy Andreev limit.  (\ref{quad_delta_1}) bosonizes to \begin{equation} \label{bosonic_delta_1} \delta_1 = \frac{P_{LL}}{2L} \int_0^L dx \left[ \sin(\phi+\theta') + \sin(\phi-\theta') \right]. \end{equation}  It is instructive to expand out (\ref{bosonic_delta_1}) in oscillator modes and verify explicitly that the bosonized expression commutes with the bosonized Hamiltonian, anti-commutes with $\delta_2$, and squares to $1$.  We represent $\delta_1$ as \be \delta_1=\left( \begin{array}{cc} 0 & \delta_1^{-+} \\ \delta_1^{+-} & 0 \end{array} \right) \ee with respect to the decomposition of $\HHlow$ into the two $\phi(0) = \pm \frac{\pi}{2}$ sectors.  Note that $\delta_1$ is purely off-diagonal because the factor of $P_{LL}$ in (\ref{bosonic_delta_1}) exchanges the two sectors.  This immediately shows that $\delta_1$ anti-commutes with \be \delta_2 = \left( \begin{array} {cc} I & 0 \\ 0 & -I \end{array} \right). \ee  The bosonized expressions for both $\delta_1^{+-}$ and $\delta_1^{-+}$ are simply (\ref{bosonic_delta_1}) with the factor of $P_{LL}$ stripped off, but it is useful to re-write them as \bea \delta_1^{+-} &=& \int_0^L dx \frac{\cos((\phi-\frac{\pi}{2})+\theta') + \cos((\phi-\frac{\pi}{2}) - \theta')}{2L} \nonumber \\ \delta_1^{-+} &=& -\int_0^L dx \frac{\cos((\phi+\frac{\pi}{2})+\theta') + \cos((\phi+\frac{\pi}{2}) - \theta')}{2L} \nonumber.\eea  This form removes the constant $\pm \frac{\pi}{2}$ from $\phi$ and in particular shows that only terms with an even number of creation/annihilation operators appear with non-zero coefficients.  Furthermore, the coefficient of a potentially energy-violating term would be the real part of an oscillatory integral which is purely imaginary; hence $\delta_1$ commutes with the Hamiltonian.  Similarly one can compute $\delta_1^2$ directly and see that it equals $1$, though it is instructive to check this explicitly on some low energy subspaces.  The lowest lying states are \bea |0\rangle &:& E=0 \nonumber \\ a_0^\dag |0\rangle &:& E=\frac{\pi v_F}{4L} \nonumber \\ \frac{1}{\sqrt{2}} (a_0^\dag)^2 |0\rangle &:& E=\frac{\pi v_F}{2L} \nonumber \\ \{ \frac{1}{\sqrt{6}} (a_0^\dag)^3 |0\rangle, a_1^\dag |0\rangle \} &:& E=\frac{3 \pi v_F}{4L} \eea  One can explicitly compute that on the lowest $3$ subspaces $\delta_1^{+-}$ acts as $+1,-1,-1$ respectively, whereas on the $E=\frac{3\pi v_F}{4L}$ subspace it acts as \be \left(\begin{array}{cc} -\frac{1}{3} & -\frac{2\sqrt{2}}{3} \\ -\frac{2\sqrt{2}}{3} & \frac{1}{3} \end{array} \right), \ee a non-trivial matrix that squares to $+1$.

When $g \neq 1$, we have \bea H &=& \frac{\pi v_F}{gL} \sum_{n\geq 0} \left[ -\frac{1}{2} \left(\hatppn\right)^2 + \frac{(g\omega_n)^2}{2} \hatpn^2 \right]. \eea  We can still expand $\phi$ in modes as in (\ref{pe}), but this time
\begin{align} \phi_n &= \frac{1}{\sqrt{2g \omega_n}} \left(a_n+a_n^\dag\right) \nonumber \\ \frac{\partial}{\partial \phi_n} &= \sqrt{\frac{g \omega_n}{2}} \left(a_n - a_n^\dag \right). \end{align}  To get rid of unwanted $g$'s we {\it define}
\be \label{p_g} \phi^g(x) = \pm \frac{\pi}{2} + \sqrt{g} \int_0^x dy \, \partial_y\phi(y). \ee
$\phi^g(x)$ is thus diagonal with respect to the $\phi(0) = \pm \pi / 2$ sector decomposition, with the choice of sign in (\ref{p_g}) corresponding to the choice of sector.  Essentially, we have used the fact that $\phi(0)$ is pinned to give well-defined meaning to the operator $\sqrt{g} \phi(x)$.  Similarly we can use the fact that $\theta'(L)=0$ to define
\be {\theta'}^g(x) = -\frac{1}{\sqrt{g}} \int_x^L dy \,\partial_x \theta'(y) \ee
The virtue of the operators $\phi^g(x), {\theta'}^g(x)$ is that their mode expansions are identical to those of $\phi(x), \theta'(x)$ in the free case, given by (\ref{phi}) and (\ref{theta}).  Thus $\delta_1^g$, defined by
\be \label{bosonic_delta_1_g} \delta_1^g = \frac{P_{LL}}{2L} \int_0^L dx \left[ \sin(\phi^g+{\theta'}^g) + \sin(\phi^g-{\theta'}^g) \right], \ee
has all the necessary properties: it anti-commutes with $\delta_2$, commutes with $H$, and squares to $1$.  It can also be written in terms of the fermions.  Let us see how that works explicitly for small $\epsilon = g-1$.  We expand to leading order:
\begin{align} \phi^g(x) &\approx \phi(x) + \frac{\epsilon}{2} \int_0^x dy\, \partial_y \phi(y) \\ {\theta'}^g(x) &\approx {\theta'}(x) + \frac{\epsilon}{2} \int_x^L dy\, \partial_y \theta'(y), \end{align}
so that
\begin{align} \label{approx_sin} \sin&(\phi^g + {\theta'}^g) \approx \sin(\phi+\theta') \nonumber \\ &+ \frac{\epsilon}{2} \cos(\phi+\theta') \left[\int_0^x dy\, \partial_y \phi(y)+\int_x^L dy\, \partial_y \theta'(y) \right]. \end{align}
We now rewrite (\ref{approx_sin}) in terms of fermions using
\begin{align} \label{ferm} \partial_x \phi &= \pi \left(r^\dag r -\ell^\dag \ell \right) \\ \partial_x \theta' &= \pi \left(r^\dag r + \ell^\dag \ell \right), \end{align}
to obtain:
\begin{align} &P_{LL} \sin(\phi^g(x) + {\theta'}^g(x)) = -\frac{i}{2} (r(x)-r^\dag(x)) \nonumber \\ &+ \frac{\pi \epsilon}{4} (r(x)+r^\dag(x))\left[\int_0^L dy\, (r^\dag r + \ell^\dag \ell) - 2 \int_0^x dy\, \ell^\dag \ell \right]. \end{align}
A subtle point is that (\ref{ferm}) has to be normal ordered with respect to the ground state of the free fermions with the tunneling term included; in practice this means replacing $r^\dag r \rightarrow (r^\dag r + r r^\dag) / 2$, etc.  With a similar expression for $P_{LL} \sin (\phi^g(x) - {\theta'}^g(x))$ we finally obtain the interacting Majorana mode
\be \delta_1^g = \delta_1 + \delta_1' \ee
with
\begin{align} \label{delta_prime} &\delta_1' = \frac{\pi \epsilon}{8L}\int_0^L dx\,(r+r^\dag-\ell-\ell^\dag) \int_0^L dy\,(r^\dag r + \ell^\dag \ell) \nonumber \\ &+ \frac{\pi \epsilon}{4L} \int_0^L dx \int_0^x dy\, \big{[} (\ell(x) + \ell^\dag(x)) r^\dag r(y) \nonumber \\ & \,\,\,\,\,\,\,\,\,\,\,\,\,\,\,\,\,\,\,\,\,\,\,\,\,\,\,\,\,\,\,\,\,\,\,\,\,\,\,\,\,\,\,\,\,\,\, - (r(x)+r^\dag(x)) \ell^\dag \ell(y) \big{]}. \end{align}
Likewise we can also expand the Hamiltonian in Fermions:
\be H = H_0 + H' \ee
with
\be H_0 = \frac{i v_F}{2} \int_0^L dx \left(-r^\dag \partial_x r - r \partial_x r^\dag + \ell^\dag \partial_x \ell + \ell \partial_x \ell^\dag \right) \ee
and
\be H' = -2\pi \epsilon v_F \int_0^L dx \, r^\dag r \ell^\dag \ell. \ee
Note that the $H_0$ we have differs from the usual one by a boundary term; again this boundary term effectively takes into account the tunneling at $x=0$.  We now need to check $[H, \delta_1^g]=0$, which at leading order reduces to checking that
\be \label{final_eq} [H_0, \delta_1'] = [\delta_1, H'] \ee
(\ref{final_eq}) can be explicitly verified with some algebra, showing that (\ref{delta_prime}) is the correct leading order interacting correction to the Majorana mode $\delta_1$.

\bibliography{topological_wires}

\end{document}